\documentclass[fleqn,usenatbib]{mnras}

\usepackage{newtxtext,newtxmath}

\usepackage[T1]{fontenc}
\usepackage{ae,aecompl}

\usepackage{graphicx}	
\usepackage{amsmath}	
\usepackage{amssymb}	

\usepackage{bm}
\usepackage{color}

\usepackage{lscape}
\usepackage{subfigure}
\usepackage{multirow}
\usepackage{url}
\usepackage{threeparttable}

\title[The origin of UFO in AGN]{UV line driven disc wind as the origin of ultrafast outflows in AGN}

\author[M. Mizumoto et al.]{
Misaki Mizumoto$^{1,2,3}$\thanks{
E-mail: mizumoto.misaki.n68@kyoto-u.jp (MM)},
Mariko Nomura$^{4}$,
Chris Done$^{1,7}$,
Ken Ohsuga$^{5}$, and
\newauthor
Hirokazu Odaka$^{6,7}$
\\
$^{1}$Centre for Extragalactic Astronomy, Department of Physics, University of Durham, South Road, Durham DH1 3LE, UK\\
$^{2}$Hakubi Center, Kyoto University, Yoshida-honmachi, Sakyo-ku, Kyoto 606-8501, Japan\\
$^{3}$Department of Astronomy, Graduate School of Science, Kyoto University, Kitashirakawa-oiwakecho, Sakyo-ku, Kyoto 606-8502, Japan\\
$^{4}$Faculty of Natural Sciences, National Institute of Technology (KOSEN), Kure College, 2-2-11 Aga-minami, Kure, Hiroshima 737-8506,\\ Japan\\
$^{5}$Center for Computational Sciences, University of Tsukuba, 1-1-1 Ten-nodai, Tsukuba 305-8577, Japan\\
$^{6}$Department of Physics, Graduate School of Science, The University of Tokyo, Hongo, Bunkyo-ku, Tokyo 113-0033, Japan\\
$^{7}$Kavil IPMU (WPI), The University of Tokyo, 5-1-5 Kashiwanoha, Kashiwa, Chiba 277-8583, Japan
}
\date{Accepted XXX. Received YYY; in original form ZZZ}

\pubyear{2019}

\begin{document}
\label{firstpage}
\pagerange{\pageref{firstpage}--\pageref{lastpage}}
\maketitle

\begin{abstract}
UltraFast Outflows (UFO) are observed in some active galactic nuclei (AGN), with blueshifted and highly ionised Fe-K absorption features.  
AGN typically have a UV bright accretion flow, so UV line driving is an obvious candidate for launching these winds. However this mechanism requires material with UV opacity, in apparent conflict with the observed high ionisation state of the wind. 
In this paper we 
synthesise the X-ray energy spectra resulting from different lines 
of sight through a state of the art radiation hydrodynamics 
UV line driven disc wind simulation. We demonstrate that there are some lines of sight which only intercept highly ionised and fast outflowing material. The cooler material required for the UV line driving acceleration is out of the line of sight, close to the disc, shielded from the X-rays by a failed wind. We fit these simulated wind spectra to data from the archetypal UFO source PG 1211+143 and show that they broadly reproduce the depth and velocity of the iron absorption lines seen. This directly demonstrates that UV line driving is a 
viable mechanism to launch even the fastest UFOs. 
We simulate microcalorimeter observations of this wind and show that
their high energy resolution can resolve the detailed structure in the wind
and recover the wind energetics when combined with 
models which correctly estimate the line formation radius of the wind.
New data from microcalorimeters will pave the way for physical predictions of AGN wind feedback in cosmological simulations.
\end{abstract}




\section{Introduction}

Strong ultrafast outflows (UFO) are observed in many active galactic nuclei (AGN), as blueshifted, highly ionised H and He like iron K absorption features (e.g.\ \citealt{kin03a,pou03,ree09}).
Systematic studies of X-ray archival data obtained by CCD detectors revealed that
UFOs are detected in about a half of AGNs across a wide range in mass and mass accretion rate;
the outflowing gas is fast ($v_{\rm w}\sim0.05c-0.3c$, where $c$ is the light velocity) and highly ionised, with large column density of $N_{\rm H}\sim10^{22}-10^{24}$~cm$^{-2}$ (e.g.~\citealt{tom11}).
The kinetic  power and momentum carried by this wind can affect the host galaxy evolution (e.g.\ \citealt{kin16}), with the 
sub-pc scale UFO coupling to and powering kpc-scale cold outflows (e.g.~\citealt{cos20} for a state of the art simulation).

However, while these winds can be tracked observationally, the physical mechanism(s) responsible for their launch and acceleration are still unknown. The 
fast velocity indicates that the material is launched close to the vicinity of the supermassive black hole (SMBH), and are only three known mechanisms which could produce this wind, namely
radiation pressure (continuum driven) for super-Eddington sources, radiation pressure (UV line driven) for sub-Eddington sources, and the Lorentz force (magnetically driven). 

Clearly the radiation force will overcome gravity when $L > L_{\rm Edd}$, where 
$L_{\rm Edd}$ is the Eddington luminosity defined from Thompson scattering on free electrons. 
However, many AGNs with UFOs are sub-Eddington, where continuum driving alone cannot accelerate the material. Instead, 
the wind can be efficiently launched even in the sub-Eddington AGNs if the radiation couples more efficiently to the material than just electron scattering. Bound-free (edges) and bound-bound (line) absorption can give opacity which is 
10--1000 times larger than
electron scattering \citep{ste90}. UV lines are especially effective, as they can absorb enough momentum to accelerate the material so that the line is shifted from resonance, so there is new continuum to absorb. Such 
UV line driven disc winds can be extremely powerful even for sub-Eddington flows
\citep{pro00,pro04,ris10,nom13,nom16},
and are also supported by some observations (e.g.\ \citealt{mat17}).
However, this mechanism requires that the material has quite low ionisation state in order that there are substantial populations of the UV absorbing ions, in apparent conflict with the observed very high ionisation state of UFOs (e.g.~\citealt{tom10,fuk15}). This leaves only the final mechanism, which is magnetic driving  (e.g.~\citealt{bla82,kon94,fuk15}), but this depends on the (currently unknown) magnetic field configuration, so this wind is difficult to determine. We do not treat it in this paper.

\citet{hag15} proposed that the disc wind geometry could 
circumvent the issue of the ionisation state of UV line driven winds. Low
ionisation material which is efficiently accelerated should be close to the 
disc, so may be out of the line of sight, and only ionised after it reaches 
the escape velocity. There is evidence for this geometry in the numerical simulations of UV line driven disc winds. These show the vertical structure of material lifted from the disc photosphere by UV line driving. At small radii, the disk photosphere is lifted
upwards by the strong UV line driving on the low ionisation material, but this is 
directly illuminated by the central X-ray source. This ionises the 
material, stopping
the acceleration before the wind reaches escape velocity so that the material falls back as a failed wind. The vertical density structure from this shields outer material from the central X-ray radiation. If this shielding is sufficiently strong then the disc wind from larger radii can reach its escape velocity before it is directly illuminated and overionised. The idea of an X-ray shielding region was first proposed by \citet{mur95}, using `hitchhiking' gas pulled up from the disc by the pressure gradient. The alternative of a failed wind was identified by \citet{pro00} as a natural feature of UV line driving in a disc geometry (see also  \citealt{pro04,ris10}). 
\citet{hag15} suggested that this could produce some lines of sight which intercept the UV line driven disc wind only after it is overionised, producing highly ionised and fast outflowing absorption features. 

Here we directly test this idea by synthesising the spectral features from Monte Carlo radiation transport through a 
radiation hydrodynamic simulation of the UV line driven disc wind. 
This was done previously by \citet{sim10b}, using the simulation of \citet{pro04}, but while that UV line driven wind simulation did show lines of sight with only highly ionised material, the outflow velocity was only
$\sim 0.05c$, rather slower than the most powerful UFO. Here we use instead a more recent UV line driven wind simulation \citep{nom20}, where the wind is faster, mainly due to differences in handling the radiation (\S\ref{sec6}).
We compare the synthesized spectra (\S\ref{sec3}) from the \citet{nom20} wind to 
{\it XMM-Newton} data of PG 1211+143 and demonstrate explicitly that this UV line  driven disc wind simulation can fit the  powerful UFO in this AGN (\S\ref{sec4}). We then simulate these models at much higher spectral resolution to show what will be possible with future microcalorimeter data from {\it the X-ray Imaging and Spectroscopy Mission} ({\it XRISM}) and {\it the Advanced Telescope for High-Energy Astrophysics} ({\it Athena}) in \S\ref{sec5}. We show how 
standard phenomenological fits to the absorption features 
generally underestimate the UFO kinetic power, and discuss its implications for AGN feedback in \S\ref{sec6}.

\section{Methods} \label{sec2}
\subsection{Radiation hydrodynamics simulation}\label{sec2.1}
\citet{nom20} simulated the UV line driven disc wind using their radiation hydrodynamic code.
We use their resulting density and velocity structure as input 
into our Monte Carlo radiation transport code. We review the hydrodynamic simulation here for completeness. 

The calculation is two-dimensional and axisymmetric about the rotation axis of the disc. 
The basic equations of the hydrodynamics are the equation of continuity, the equations of motion, and the energy equation (including Compton heating/cooling, X-ray photoionisation heating, recombination cooling, bremsstrahlung cooling, and line cooling).

The radiation force per unit mass, ${\bm f}_{\rm rad}$, is 
\begin{equation}
{\bm f}_{\rm rad}=\frac{\sigma_{\rm e}{\bm F}_{\rm D}}{c}+\frac{\sigma_{\rm e}{\bm F}_{\rm line}}{c}M,
\end{equation}
where $\sigma_{\rm e}$ is the mass-scattering coefficient for free electrons,  ${\bm F}_{\rm D}$ is the radiation flux of the total accretion flow emission, whereas
${\bm F}_{\rm line}$ is the radiation flux from the disc  
integrated across the UV transition band ($200-3200$\AA), and
$M$ is the force multiplier as a function of ionisation parameter, density, and velocity gradient along the line of sight (\citealt{ste90}; recalculated in \citealt{nom17}). 
The ionisation parameter, $\xi$, is assumed to be set only by the X-ray irradiating source with luminosity $L_{\rm X}$. This is 
assumed to be a point source at the centre, so that $\xi=4\pi F_{\rm x}/n$, where $F_{\rm X}=e^{-\tau_{\rm X}} L_{\rm X}/(4\pi r^2)$\footnote{In this paper $r=\sqrt{x^2+y^2+z^2}$ in the Cartesian coordinates, whereas $R=\sqrt{x^2+y^2}$.}, takes into account the shielding effect of gas along the line of sight.
Here $\tau_{\rm X}=\int_0^r \sigma_{\rm X} n dr $ and $\sigma_{\rm X}=\sigma_{\rm e}$ for $\xi\geq10^5$ or $100\sigma_{\rm e}$ for $\xi<10^5$. This step function approximately incorporates the increase in opacity due to the increasing numbers of bound-free transitions in lower ionisation material \citep{pro02}. 

However, in the most explored radiation hydrodynamic simulation in \citet{pro04}, they used $\sigma_{\rm X}=\sigma_{\rm e}$ for all $\xi$. This gives the minimum possible X-ray shielding by the failed wind. They also assume $\sigma_{\rm UV}=0$, i.e. that $\bf F_{\rm line}$ is unattenuated as the velocity structure of the wind Doppler shifts the line transition into unabsorbed UV continuum. However, electron scattering is clearly a lower limit on the attenuation of the UV flux, so \citet{nom20} also include $\sigma_{\rm UV}=\sigma_{\rm e}$ assuming the same line of sight as for the X-rays. 

Another, more physical, difference is that these 
winds are often so powerful that the mass loss rate in the outflow is comparable to (or even larger than) the mass accretion rate through the disc which powers the outflow. 
Hence \citet{nom20} also includes the reduction in mass accretion rate (and consequent UV luminosity) through the disc caused by the wind (see also \citealt{dav11}). 
The code makes the approximation that the total mass loss rate in the wind, $\dot{M}_{\rm out}$, comes from a single wind launching radius $R_{\rm launch}$. 
Thus the mass accretion rate for $R>R_{\rm launch}$ is original mass accretion rate supplied through the outer disc, $\dot{M}_{\rm sup}$, but it drops to $\dot{M}_{\rm BH} = \dot{M}_{\rm sup}-\dot{M}_{\rm out}$ for $R<R_{\rm launch}$. The wind also carries away the specific angular momentum from Keplerian rotation at its wind launching radius. The sum over all angles of the angular momentum and mass outflow rate in the wind is then used to define the single wind launching radius so that both mass and angular momentum are conserved \citep{nom20}. 

The temperature and density distribution on the $\theta=\pi/2$ plane is estimated using the standard disc model for both the inner and outer region using their different mass accretion rates. 
The radiation fluxes (${\bm F}_{\rm D}$ and ${\bm F}_{\rm line}$) are calculated based on this temperature distribution.
The hydrodynamics are calculated over a computational domain from 
$r=60-3000\,R_g$ in the northern hemisphere, where $R_g$ is the gravitation radius.
Radiation from the disc and X-ray source from within $60\,R_g$ are included but its wind is not calculated as it is assumed to be  extremely ionised in this region. 
The rotational (=azimuthal) velocity is set by the equilibrium between gravitational and centrifugal force. The typical rotational velocity is 0.005c--0.05c, which is smaller than the typical radial velocity (0.1c--0.25c) by $\sim$one magnitude.

As a result of this calculation, we get the two dimensional map of the wind density and velocity as a function of time.
We choose the simulation where $M_{\rm BH}=10^8\,M_\odot$ and $\dot{M}_{\rm sup}=0.9\dot{M}_{\rm Edd}$,
where $\dot{M}_{\rm Edd}=L_{\rm Edd}/(\eta c^2)$ with the energy conversion efficiency $\eta=0.06$ (spin 0). A powerful wind is launched from a radius of 
$R_{\rm launch}=120\,R_g$, with $\dot{M}_{\rm out}=0.56\dot{M}_{\rm Edd}$
so that only $\dot{M}_{\rm BH}=0.34\dot{M}_{\rm Edd}$ accretes through the inner disc onto the black hole. 
The X-ray ionising luminosity is assumed to be 10\% of the mass accretion power onto the black hole, i.e., $L_{\rm X}=0.1\times0.34L_{\rm Edd}=0.034L_{\rm Edd}$ in the converged solution. 
The wind structure is time dependent even after the initial transient structures subside.
We study three snapshots of the wind after it has reached steady state,
and show detailed results for the one which best matches the observational features
seen in PG 1211+143. The momentum flux and kinetic power are calculated at the outer boundary of the simulation box, where the wind has already reached its terminal velocity, as $\dot{P}_{\rm w}=1.7(L_{\rm Edd}/c)$ and $L_{\rm w}=0.17L_{\rm Edd}$. The other snapshots 
are shown in Appendix \ref{sec:a1}. 

\begin{figure}
\centering
\includegraphics[width=3.7in]{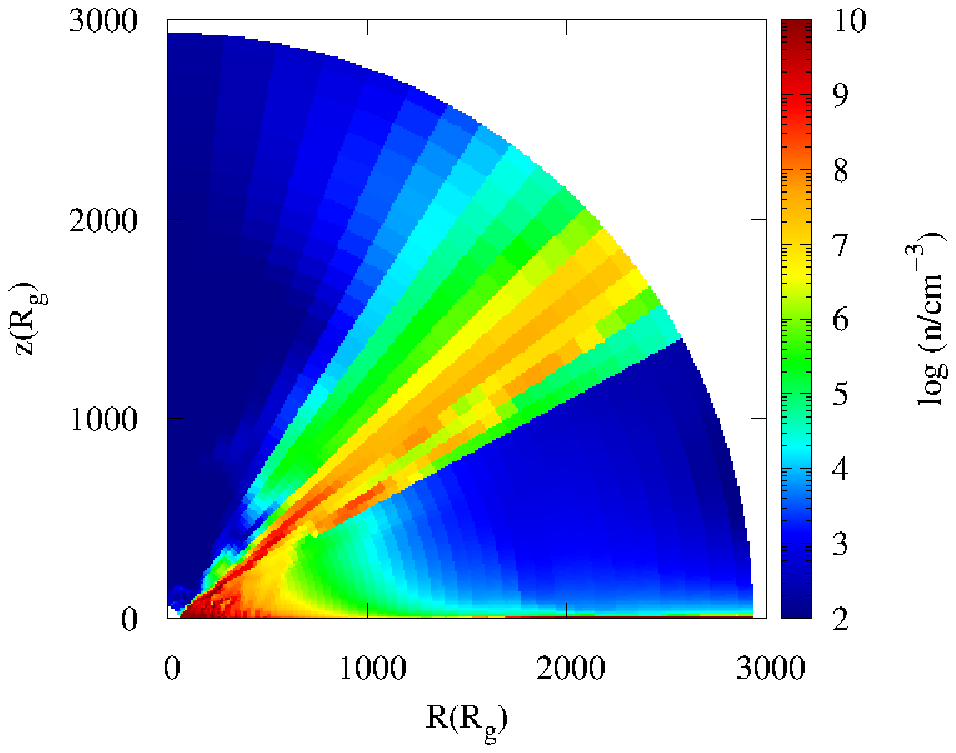}
\includegraphics[width=3.7in]{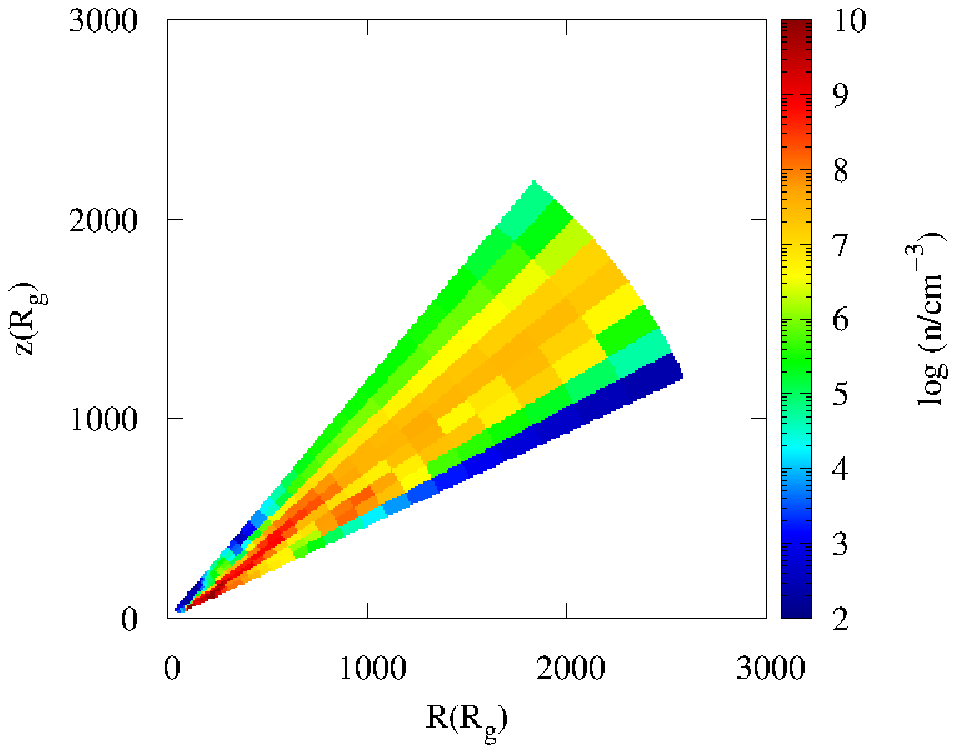}
\caption{Density structure of the UV line driven disc wind.
The upper shows the original grid obtained from the radiation hydrodynamics simulation \citep{nom20},
whereas the lower is the resampled one.
}
\label{fig:wind_density}
\end{figure}

We re-sample the hydrodynamic simulation result on a sparser grid in order to make the radiation transport calculation more tractable. We consider material only in the wind region, 
defined as $\theta=40$~deg to $65$~deg, and convert it into 36 (radial) $\times$ 9 (polar) cells. 
The upper panel of Figure \ref{fig:wind_density} shows the density structure from the original hydrodynamic calculation, while the resampled sparser grid used for the radiation transport is shown in the lower panel.

\subsection{Calculation of the ionisation state}
\label{sec2.2}

\begin{figure}
\centering
\includegraphics[width=4in]{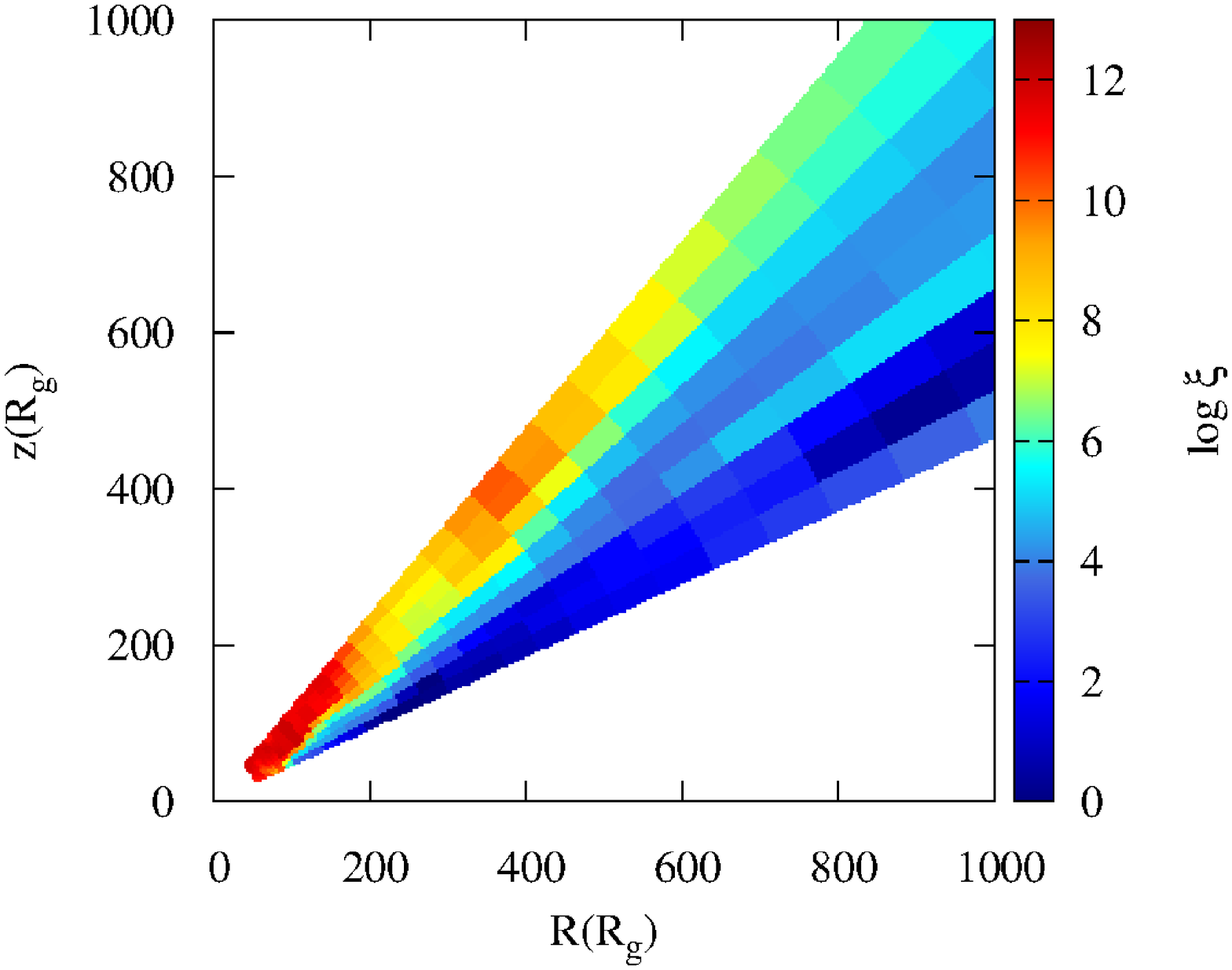}
\includegraphics[width=4in]{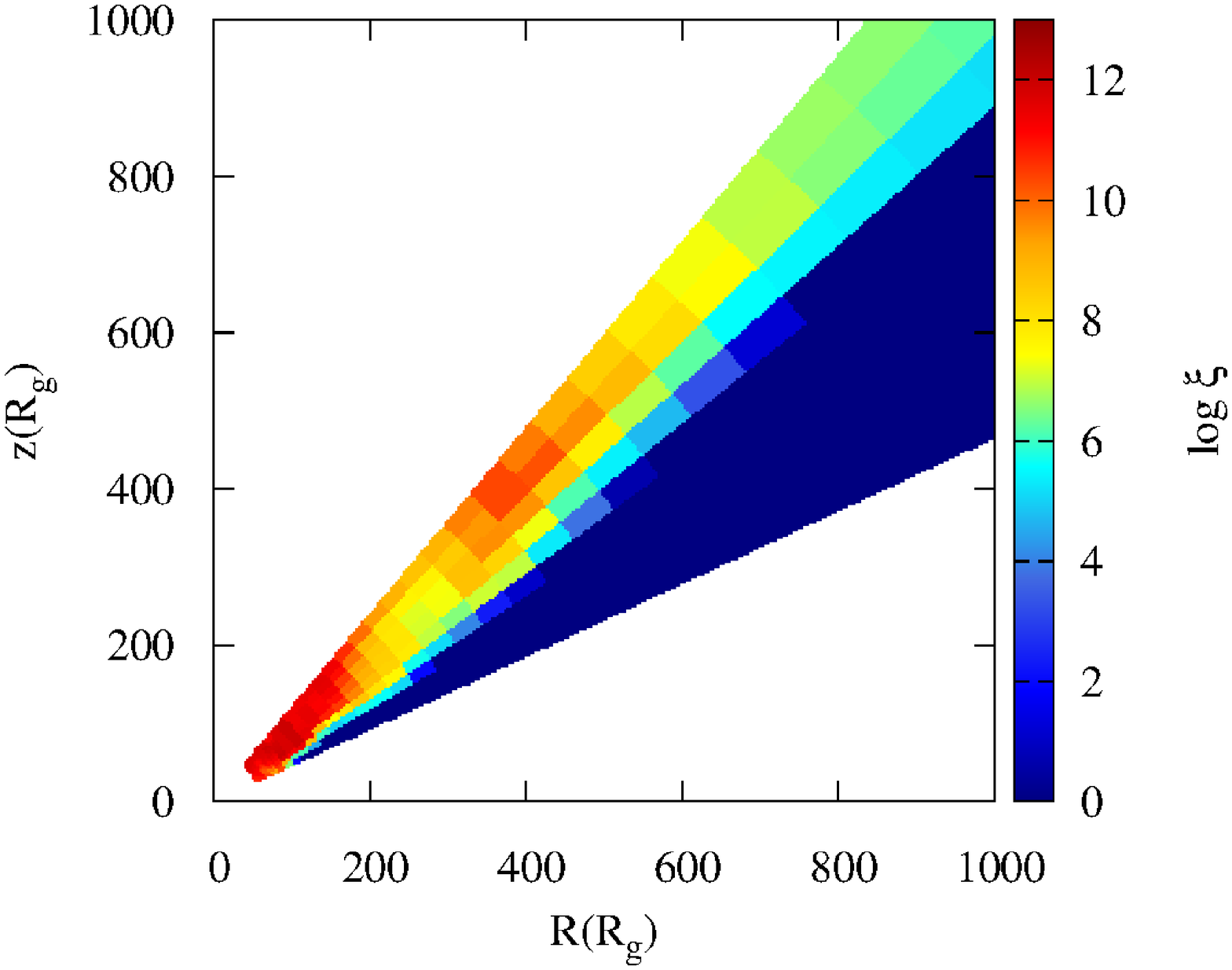}
\caption{
Upper panels shows the 
ionisation parameter map used in the {\sc monaco} radiation transfer, recalculated from the 
original simulation using  {\sc xstar}. The lower panel shows the ionisation parameter map from the 
original simulation of \citealt{nom20}). Both share the same colour mapping. The underside of the wind (and its base) are slightly less ionised assuming full opacity than predicted from the step function, but the upper part of the wind where the UFO structures are seen are very similar in ionisation state.
We note that the region with $\log\xi<0$ has the same colour as that with $\log\xi=0$, both of which have only the neutral Fe atoms.
}
\label{fig:wind_logxi2}
\end{figure}

The radiation hydrodynamic code used a step function approximation for the X-ray opacity in order to calculate the effective shielding of the failed wind, so the ionisation state within the wind is quite approximate. Therefore we recalculate the ion populations in each grid point of
the gas using the full X-ray opacity as determined via {\sc xstar} (version 2.54; \citealt{kal04}) for each polar angle. The central source is assumed to emit a power law spectrum with the photon index $\Gamma=2.2$ and the ionising luminosity $L_{\rm X} = 0.034L_{\rm Edd}$.
The mass density is converted into the number density as $n=\rho/(\mu m_{\rm p})$, where $\mu=0.5$ is the mean molecular weight under the assumption that the outflowing gas is almost fully ionised.
When the outflowing velocity is high, the X-ray luminosity in the rest frame of the wind is reduced by Doppler (de-)boosting.
Therefore, we use
$L_{\rm obs}=(\gamma(1-\beta))^{-(3+\alpha)}L_{\rm X}$ as the ionising luminosity in each cell,
where $\beta=v/c$ ($<0$ for the outflowing case), $\gamma$ is the Lorentz factor, and the energy spectral index $\alpha=\Gamma-1=1.2$ (see \citealt{sch07}).
The output spectrum of each {\sc xstar} run is used for the input spectrum in the next cell. 
The turbulent velocity is fixed as $1000$~km~s$^{-1}$.
Atomic abundances are assumed to be equal to the solar abundances for all elements.
As a result of this calculation, we get the ion population and electron temperature in each grid assuming that the directly transmitted radiation is the main ionising source (Figure \ref{fig:wind_logxi2}, but see \citealt{hig14} for a discussion of scattering).

\subsection{Monte Carlo radiation transfer simulation}
We use {\sc monaco} (MONte carlo simulation for Astrophysics and COsmology, ver 1.1.3, \citealt{oda11}) for the Monte Carlo radiative transfer calculation;
the density and velocity results of the radiation hydrodynamics simulation (Figure \ref{fig:wind_density}) are input into it.
{\sc monaco} is a general-purpose framework for synthesising X-ray radiation from structured astrophysical objects, treating photon transport and photon interactions with matter. 
It uses the {\sc geant4} toolkit library \citep{geant4} for photon tracking in an arbitrary three-dimensional geometry, but has its own modules handling photon interactions.
In this code, photons are generated at a illuminating source, which is at the centre in this work, and then tracked in the system until they escape from the system toward the observers. In a photoionised plasma, the code treats the photon interactions via photoionisation, photoexcitation, and Compton scattering. It also calculates line and continuum emissions associated with radiative recombination and atomic de-excitation. Special relativity effects by the motion of the plasma are taken into account, and thus the Doppler shifts and broadening of both absorption and emission lines can be compared with observational data. General relativity is ignored since the wind size scale is sufficiently large.

This X-ray spectral calculation based on radiative transfer is static, namely the photon tracking is performed within a snapshot of a wind structure produced by the radiation hydrodynamic simulation and the ionisation state calculation. The wind structure, which is described by the density, velocity, ion populations and electron temperature at each calculation mesh, is input to the radiative transfer code. The initial photons generated with the power law spectrum at the central source, and shot outward radially with an isotropic luminosity.

The physical models of the photon interactions considered within {\sc monaco} in this work are divided into two categories: highly ionized ions and lower ions. We consider all major transitions of H- and He-like ions of C, N, O, Ne, Mg, Si, S, Ar, Ca, Fe, and Ni, whereas the lower ionisation states after Li-like are treated as neutral. This neutral treatment means that these low charge ions do not have emissions with recombination, but affect the observed spectra via continuum absorption and its following fluorescence line emissions. These neutral gas interactions are considered by \citet{oda11}. On the other hand, the physical model for the highly ionised gas was described by the previous work \citep{hag15}, and its original implementation was introduced in the context of the photoionised stellar wind in an X-ray binary \citep{wat06}. 
In the simulation code, an absorption and emissions are invoked as a sequential process. Line emissions after a photoexcitation are due to atomic cascade de-excitations with specific transition rates. Similarly, line and continuum emissions associated with a radiative recombination is followed by a photoionisation, considering the detailed balance between the ionisation and recombination, which is consistent with the photoiosation equilibrium. We use Flexible Atomic Code \citep{gu08} to calculate all atomic properties required for the calculation of the photon interactions with the H- and He-like ions, which include the list of atomic energy levels, photoionisation cross sections for destination levels, photoexcitation cross sections, atomic transition rates between bound levels, and autoionisation rates of He-like ions.

The 3D wind geometry is assumed to be axisymmetric about the rotation axis, and each torus (single cell in the 2D map) is split into 
48 (azimuthal) $\times$ 2 (polar).
As a result, the whole calculation grid is 36 (radial) $\times$ 48 (azimuthal) $\times$ 18 (polar).
We input $4\times10^7$ photons.
We created the model spectra accumulated over 10 viewing angles; face-on ($0^\circ<\theta<18.2^\circ$), angle1 ($40^\circ<\theta<42.78^\circ$), 
angle2 ($42.78^\circ<\theta<45.56^\circ$), angle3 ($45.56^\circ<\theta<48.33^\circ$), 
angle4 ($48.33^\circ<\theta<51.11^\circ$), angle5 ($51.11^\circ<\theta<53.89^\circ$), 
angle6 ($53.89^\circ<\theta<56.67^\circ$), angle7 ($56.67^\circ<\theta<59.44^\circ$), 
angle8 ($59.44^\circ<\theta<62.22^\circ$), and angle9 ($62.22^\circ<\theta<65^\circ$)
from low to high inclination.

The radial acceleration means that the
velocity change from cell to cell, $\Delta v_{\rm w}$,  can be large. We approximate this
velocity shear across each cell as $v_{\rm turb,shear}=\Delta v_{\rm w}/\sqrt{12}$ (see equation 1 in \citealt{sch07}). We assume a baseline minimum true
turbulence of $v_{\rm turb,int}=1000$ km~s$^{-1}$ so that 
each cell in the {\sc monaco} simulation
has a 
pseudo-turbulent velocity of $v_{\rm turb,ps}={\rm max}\{v_{\rm turb,int},v_{\rm turb,shear}\}$.
See Appendix \ref{sec:a2} for a demonstration of the validity of this assumption.

\subsection{Novelty and limitations of our simulation}\label{sec2.4}

Post-processing radiation transfer of a UV line driven disc wind has been done before, using the numerical data of very influential radiation hydrodynamic simulations of \cite{pro04}. Single sight lines through this density/velocity structure were calculated by \cite{sch09}, while \cite{sim10b} used a Monte-Carlo radiation transfer approach similar to that described above. These both showed that there are lines of sight
which are dominated by highly ionised material, rather than the UV opaque material required for UV line driving which is 
instead out of these sightlines, closer to the disc. However, the
velocity of the outflow in this simulation is too slow to match to the most powerful UFOs \citep{sch09,sim10b,hig14,wat18}, being 
typically $\sim 0.05c$, which is only barely in the UFO definition. Thus the results of the radiation transfer through the \cite{pro04} simulation could not be matched to the data 
from the most powerful and convincing UFO sources. 
Instead, we use the numerical data of newer simulation of \cite{nom20} as the basis for our post-processing radiation transfer. This has a much faster wind, with $v\sim 0.2c$, so we can directly compare this to the AGN data with the strong UFO. 

The change in opacities between \citet{nom20} and \citet{pro04} is likely to have a large effect on wind properties (see Section 3.2 and Appendix in
\citealt{nom20}). 
The wind is sensitive to the X-ray illumination as this sets the ionisation state of the material. The \cite{pro04} simulation has only the absolute minimum X-ray opacity along the line of sight i.e. electron scattering with cross-section $\sigma_{\rm X}=\sigma_{\rm e}$. This strongly underestimates the attenuation
of softer X-rays in partially ionised material. Instead, the \cite{nom20} simulation uses a step function approximation, with $\sigma_{\rm X}=100\sigma_{\rm e}$ for partially ionised material (see also \citealt{pro00}). This
increases the shielding of the X-ray flux, so decreases the ionisation. This increases the number of atomic transitions present, which increases the force multiplier 
and gives a more powerful, and faster wind across a wide range of AGN with different mass and mass accretion rates \citep{nom17}. 

The difference in UV opacity should also impact the wind structure, as $\sigma_{\rm UV}=0$ in \citet{pro04} clearly overestimates the UV flux on the outer disc as the wind is optically thick to electron scattering close to its base. 
\citet{nom20} assessed the impact of these assumptions and shows that the averaged outflow velocity using $\sigma_{\rm X}=\sigma_{\rm e}$ and  $\sigma_{\rm UV}=0$ \citep{pro04} is $\sim2$ times smaller than that obtained by their current setup ($\sigma_{\rm X}=\sigma_{\rm e}$, switching to $100\sigma_{\rm e}$ at $\log\xi=5$ and $\sigma_{\rm UV}=\sigma_{\rm e}$, see Figure 5 in \citealt{nom20}). Additionally, we can get more physical insight by 
ignoring pressure forces, so that the wind streamlines can be calculated using only ballistic trajectories (the {\sc qwind} approach: \citealt{que20}). 
Figure 10 in \citet{que20} shows explicitly the 
large change in wind structure with the different assumptions about the  direct flux radiation transfer between \cite{pro04} and \cite{nom20}. The wind is clearly much denser with the \citet{pro04} assumptions, but it is also much slower (Quera-Borfarull, private communication). 

While the step function approach to $\sigma_{\rm X}$ is better at capturing the changing opacity of partially ionised material, it is still clearly an approximation. We recalculate the ion populations in the wind using {\sc xstar} (see above), giving some differences between those used in our radiation transfer and the radiation hydrodynamic code. 
Figure \ref{fig:wind_logxi2} shows the extent of these differences. The upper panel shows that calculated using the full opacity from {\sc xstar} while the lower panel shows the ionisation parameter 
used in the radiation hydrodynamic code. The underside of the wind is slightly less ionised with a full treatment of the opacity, but the upper side (where the UFO absorption lines will appear) share almost the same ionisation parameters. 

Another limitation in our simulation is that we do not include the effect of scattered radiation on the wind structure. 
\citet{hig14} showed that this can strongly suppress UV line driving by raising the ionisation parameter of material which is shielded from direct X-ray flux. 
This remains a significant uncertainty and 
the steady state structures which result from this are not yet fully understood, though we note that the
{\tt qwind} code  has the potential to include this.

The  \cite{nom20} simulation does iterate to calculate the effect of the wind mass loss rate on the disc structure, 
reducing the mass accretion rate through the inner disc in response to the outflow mass loss. This makes the inner disc less bright in UV by factor of $\sim3$, while weakening the wind relative to previous simulations with constant accretion rate throughout the disc.

\section{Results}\label{sec3}

\begin{figure}
\centering
\includegraphics[width=3in]{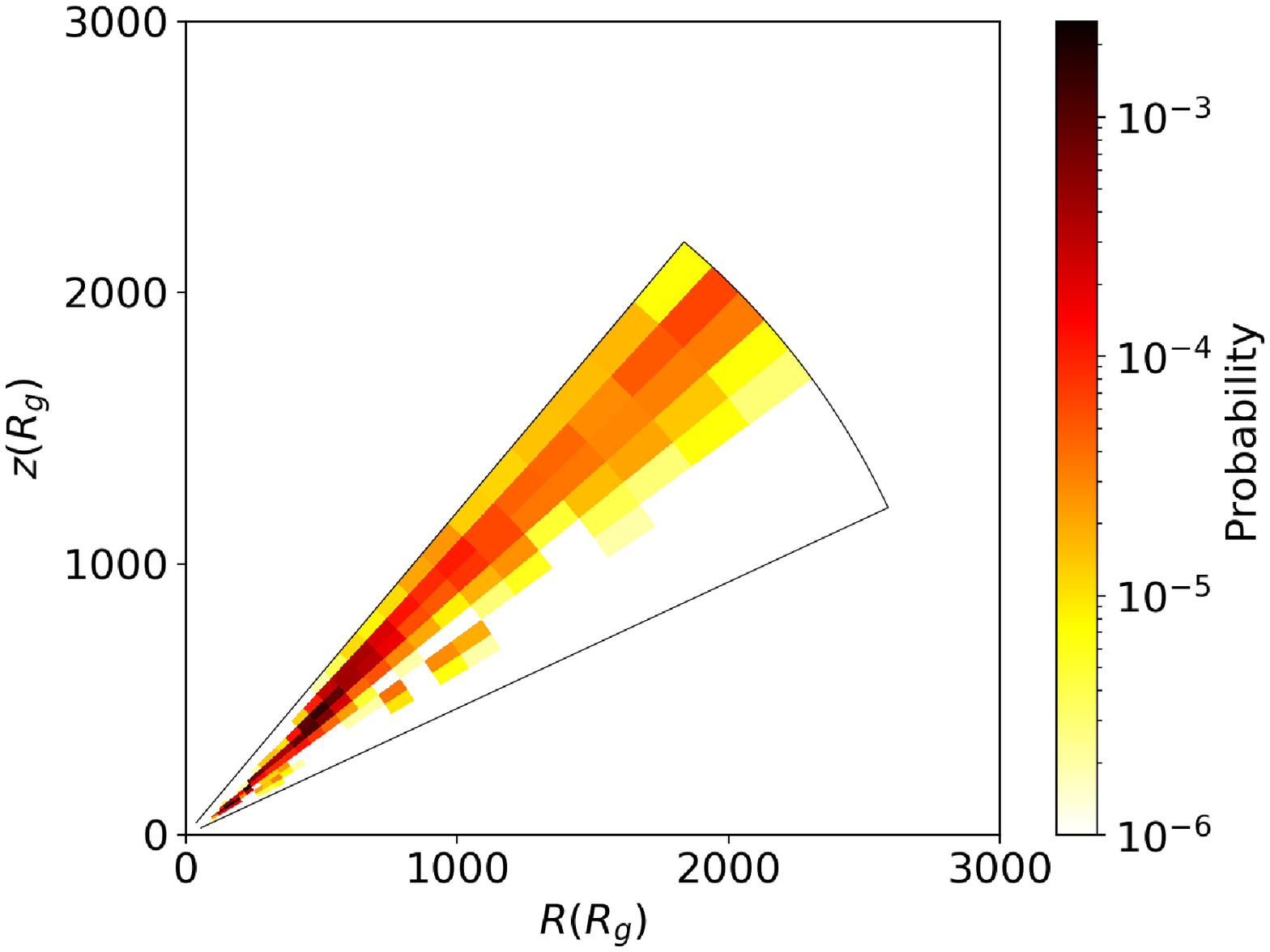}
\includegraphics[width=3in]{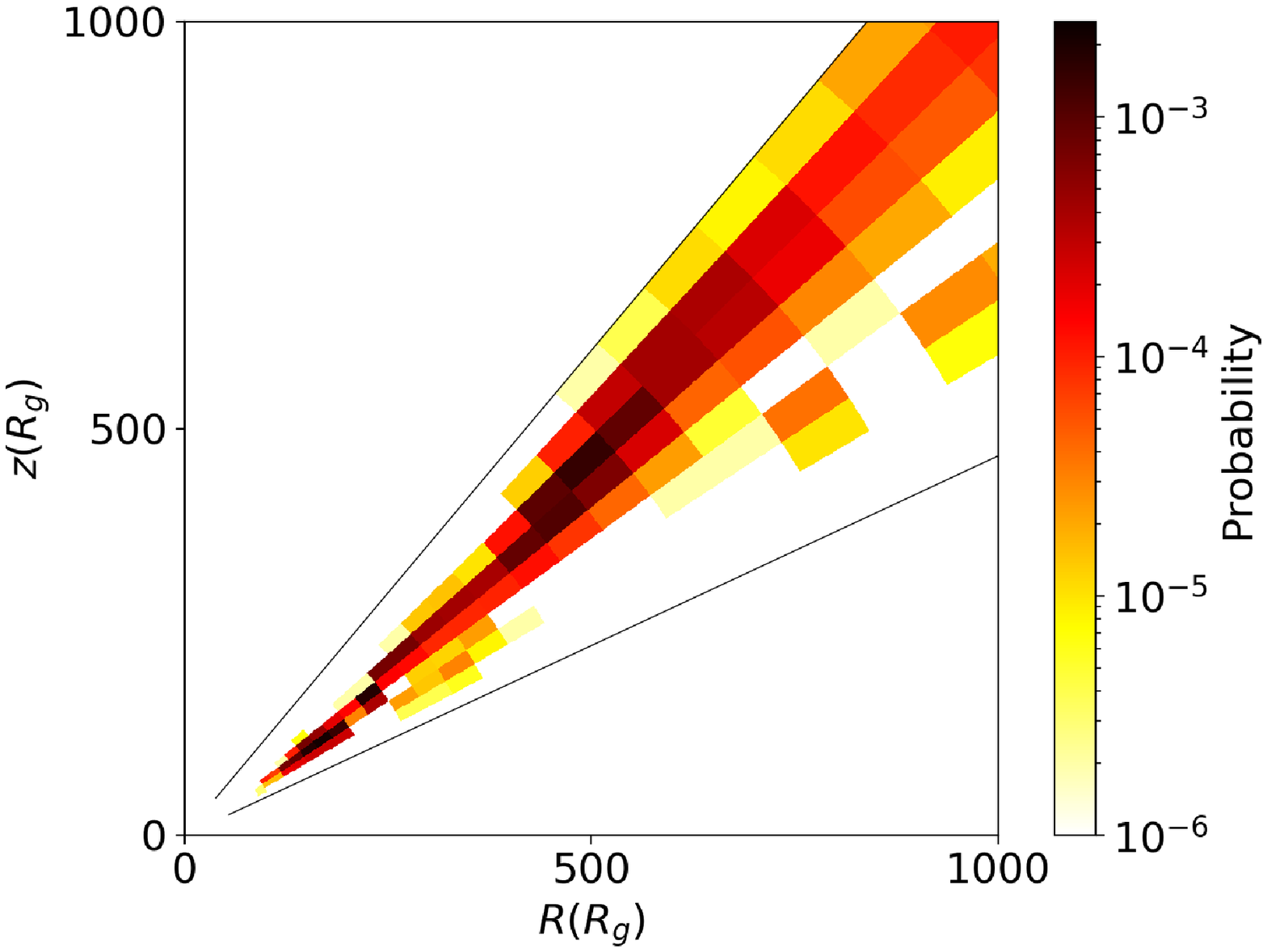}
\caption{
Probability that Fe ions absorb the input photons in each cell.
The lower panel is a closer view of the upper one.
The absorption line features will be mainly created in the dark colour region, at $(R,z)\sim(500R_g,500R_g)$.
Strong X-ray shielding exists near the disc, resulting into the blank space in the high inclination region.}
\label{fig:Feabs}
\end{figure}

\begin{figure}
\centering
\includegraphics[width=2.3in,angle=270]{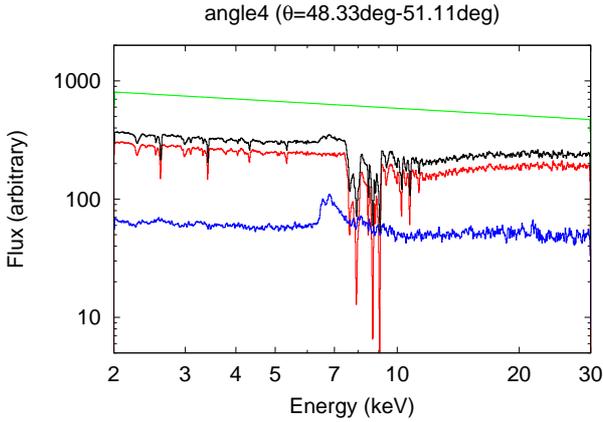}
\caption{Synthesised X-ray spectra in the angle4 case.
The vertical axis shows the arbitrary keV$^2$ flux.
The red and blue lines show the primary and scattered spectra, respectively, and the black line is the total spectrum.
The input power-law spectrum is shown in green. 
}
\label{fig:spectrum_theta4}
\end{figure}

The Monte Carlo calculation shows explicitly where 
absorption and scattering occur.
Figure \ref{fig:Feabs} shows the positions where  Fe line absorption takes place. 
This is mainly at radius $r\sim500-1000\,R_g$ along the mid-viewing angle (angle4). 
At higher viewing angles, 
most of the intrinsic X-ray flux is absorbed by 
the strong X-ray shielding and thus little line absorption occurs.
Figure \ref{fig:spectrum_theta4} shows the
output spectrum along this angle4 sightline. 
The transmitted spectrum (red), whose photons have experienced no interaction, has lower continuum than the input spectrum due to electron scattering removing photons from the line of sight, as well as showing a forest of absorption lines.
There is very little soft X-ray absorption in either the transmitted or total flux,
despite our approximation that all ions lower than 
He-like are regarded as neutral which overestimates the soft absorption. This
shows how little of the low ionised X-ray shielding gas exists along this line of sight.

The scattered spectrum (blue, including all photons which  scatter at least once), has a broad emission line at $6-7$~keV.
This emission line is mainly due to fluorescence and/or resonance scattering in the failed wind region. Its 
width is set by the range of 
Doppler shifts from different azimuths and radii \citep{miz19a}. The scattered emission has much less ionised absorption imprinted on it, contrasting to the much stronger features which can be seen in the transmitted spectrum.

The scattered continuum dilutes the 
observed absorption line depth in the total spectrum (black).
The equivalent width (EW) of the absorption line in the total spectrum will be
${\rm EW_{total}}={\rm EW_{prim}}\,f_{\rm prim}/(f_{\rm prim}+f_{\rm scat})$, where $f_{\rm prim,scat}$ is the fraction of the primary and scattered component to all the flux, respectively (i.e., $f_{\rm prim}+f_{\rm scat}=1$).
Since we can only observe the total spectrum, 
we will underestimate the line depth and hence underestimate the amount of outflowing material in objects where scattering is important.

\begin{figure}
\centering
\includegraphics[width=3.8in,angle=270]{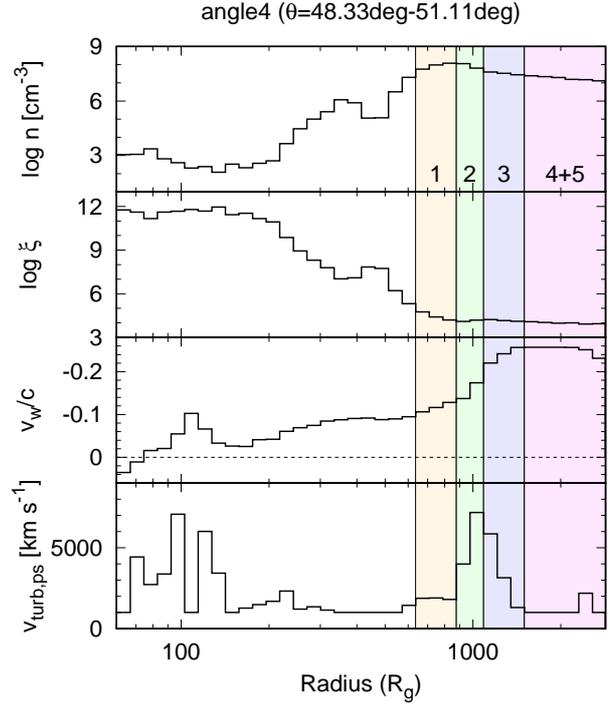}
\caption{Radius dependence of the wind parameters (density, ionisation parameter, velocity, pseudo-turbulent velocity) along angle4.
The shaded regions potentially produce the UFO absorption lines. Each colour corresponds to each absorber component listed in Figure \ref{fig:smooth} and Table\ \ref{tab:smooth}.
}
\label{fig:tau_a4}
\end{figure}

We investigate the physical properties of the wind along angle4 in Figure \ref{fig:tau_a4}.
The top two panels show that the wind density along this sightline gradually increases and reach $n\sim10^8$~cm$^{-2}$ at $r\sim600\,R_g$, at which point its ionisation state drops below $\log\xi\sim5$ so that the column density of H and He-like iron increases, producing the UFO features. However, this material is launched from the disc, where even higher column densities of the wind material make an effectively shield so that UV line driving can take place. 
This strong acceleration is also seen in the upper, 
UFO producing region ($\log\xi<5$), where the wind speed rapidly increases from $\sim0.1c$ ($r\sim600R_g$) to its terminal velocity of $\sim0.25c$ ($r\sim1300R_g$)
(see the third and fourth panels of Figure \ref{fig:tau_a4}). 
This velocity structure will produce the separate sets of UFO absorption lines in the X-ray energy spectrum.

\begin{figure}
\centering
\includegraphics[width=3.2in,angle=270]{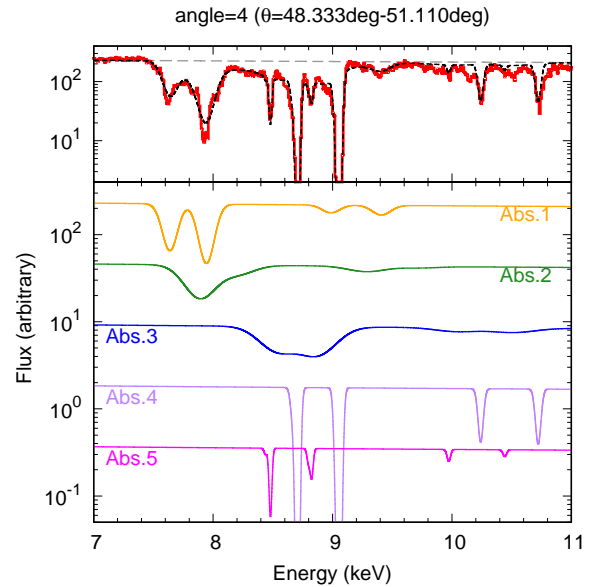}
\caption{
(Upper) The primary spectrum (red, same as Figure \ref{fig:spectrum_theta4}) is fitted by the power law continuum with $\Gamma=2.2$. (gray dashed) and five photoionised absorbers (black dotted).
(Lower) Absorption features produced by each photoionised absorber introduced.
Each absorber has the iron \ion{Ly}{$\alpha$,$\beta$} and \ion{He}{$\alpha$,$\beta$} lines, which is modelled by the {\tt kabs} model (Tomaru et al.\ in prep.).
The free parameters are the number of H-like and He-like Fe ions ($N_{\rm atom}$), velocity dispersion ($\sigma_{\rm v}$), and wind velocity.}
\label{fig:smooth}
\end{figure}

\begin{table*}
\centering
\caption{Parameters of each absorber group introduced in Figure \ref{fig:smooth}}
\label{tab:smooth}
\begin{tabular}{ccccccc}
\hline
\hline
Group & \multicolumn{2}{c}{$N_{\rm atom}$ ($\times10^{18}$~cm$^{-2}$)} & $\sigma_{\rm v}$ (km s$^{-1}$) & $v_{\rm w}$ & $N_{\rm H}$ ($\times10^{23}$cm$^{-2}$) & $\log\xi$\\ 
&  H-like & He-like &&&&\\
 \hline
Abs1 & 5.0 & 2.0 & 3400 & $-0.123c$ & 4.9 & 4.3\\  
Abs2 & 1.5 & 2.8 & 7000 & $-0.151c$ & 2.0 & 4.0 \\
Abs3 & 5.0 & 2.1 & 7400 & $-0.215c$ & 9.7 & 4.3\\
Abs4 & 7.5 & 3.5 & 900 & $-0.230c$ & 5.8 & 4.3 \\
Abs5 & 0.6 & 0.5 & 600 & $-0.210c$ & 0.5 & 4.2 \\
\hline
\end{tabular}
\end{table*}

The upper panel of Figure \ref{fig:smooth} shows a
zoom in of the region from 7--11~keV in Figure \ref{fig:spectrum_theta4},
where the iron K absorption features dominate. 
Only the primary component is shown to highlight the complex absorption. We fit the simulated data (red, same as shown in Fig 4)
using the {\tt kabs} model in {\tt xspec}
(\citealt{ued04}, updated by Tomaru et al.\ in prep).
This model uses the full Voigt profile for each line, 
calculating the line depth including all doublet structures
from the oscillator strength, turbulent velocity and column density of each atom ($N_{\rm atom}$). For each ion 
we include the K$\alpha$ and K$\beta$ transitions, tying their column density and turbulent width together.

The simulation results are not well fit by a single density/velocity component. There are clearly at least two different velocity components which give
both He- and H-like K$\alpha$ lines, one set at 7.6 and 7.95~keV
and another at 8.7 and 9.05~keV. These velocities correspond to material in regions 1 and 4 of Figure \ref{fig:tau_a4}. Although there are several radial bins within each of these regions, the turbulent width of the line means these are blended together, forming a feature which can be modelled with a single {\tt kabs} component, where the real acceleration within each region is described as an effective turbulence. 
However, even after adding this second component there are significant broad residuals. The velocity structure shown in Figure \ref{fig:tau_a4} makes it clear that there is a significant amount of material which is in a fast acceleration zone, producing a broader absorption component.

To produce a good match to this simulation data requires 5 {\tt kabs} components (Figure \ref{fig:smooth}), with parameters listed in Table \ref{tab:smooth}.
We give the equivalent Hydrogen column density ($N_{\rm H}$) of each iron ion, and we use the ratio of these to derive
$\log\xi$ assuming that the AGN spectral energy distribution (SED) has a hard extreme-UV ionising flux with an X-ray to optical luminosity ratio ($\alpha_{\rm OX}$) of 1.5 (see Figures 3 and 5 in \citealt{kri18}). 
The estimated ionisation parameters are around $\log\xi\sim4-4.3$, which is the 'habitable zone' for the ionised Fe absorption lines.

We identify each {\tt kabs} absorber with the radial regions identified in Figure \ref{fig:tau_a4}.
At $r\sim600-900\,R_g$ (Region 1 in Figure \ref{fig:tau_a4}), the wind radial velocity is $\sim-0.12c$ and its velocity shear is not large, giving relatively narrow features. 
This region has a density peak along the line of sight, and thus makes the (relatively) slow, narrow, and strong  \ion{Ly}{$\alpha$} and  \ion{He}{$\alpha$} absorption lines in 7.5--8.1~keV, as well as the \ion{Ly}{$\beta$} line at 9.4~keV.
Next, the wind velocity dramatically increases at $r\sim900-1300\,R_g$, which results in much broader line features (Regions 2 and 3). 
We note that the column density of Abs2 and 3 might be overestimated because constraining the EW of broad lines is difficult in general.
Last, the wind reaches its terminal velocity (Region 4+5). The velocity shear is again small so these lines are relatively narrow, but very fast, at 8.4--9.1~keV. 
This region shows the limitations of our approximation using turbulence as a proxy for velocity shear. 
Obviously the slight but continuous 
wind deceleration from $-0.23c$ to $-0.21c$ and slight but continuous drop in density 
across this region should produce a single broad absorption line, whereas in our approximation the lines are separated by more than the pseudo-turbulence width, so appear as discrete features (Abs4 and 5).
We also note that the column density in Abs5 is smaller than that in Abs4 by one magnitude, and thus the spiky feature due to Abs5 may be ignorable in the `real' wind.
Hereafter we call Abs1+2 as a slower wind, whereas Abs3--5 is the faster wind. See Appendix \ref{sec:a2} for a fuller discussion of the physical reality of these two main absorption systems. 

The wind becomes faster with larger $r$ due to the acceleration associated with UV line driving, so the slower wind 
locates closer to the black hole than the faster wind.
This is opposite to the standard assumption that 
the line formation radius is comparable to the wind launching radius and that 
the escape velocity from the wind launching radius is approximately equal to the wind velocity: this predicts that the wind 
becomes slower with larger $r$
(e.g.~\citealt{tom11,gof13}).

We find that most of the absorption features seen in our simulation can be explained with the multi-velocity components.
Thus the detection of multi-velocity components in the data does not necessarily mean that there are several different types of wind present in the system. 
Any wind driven by radiation pressure is subject to instabilities formed by shadows, producing a complex velocity/density structure from a single wind launching mechanism. 
We also stress that the large velocity dispersion ($\sigma_{\rm v}$) is created due to change of radial velocity of the wind; it is not because of the turbulent velocity.
In other words, even when the intrinsic turbulent velocity is small (1000~km~s$^{-1}$), the observed line can be much broader.

\begin{figure*}
\begin{center}
\subfigure{
\resizebox{5.7cm}{!}{\includegraphics[width=3.2in,angle=270]{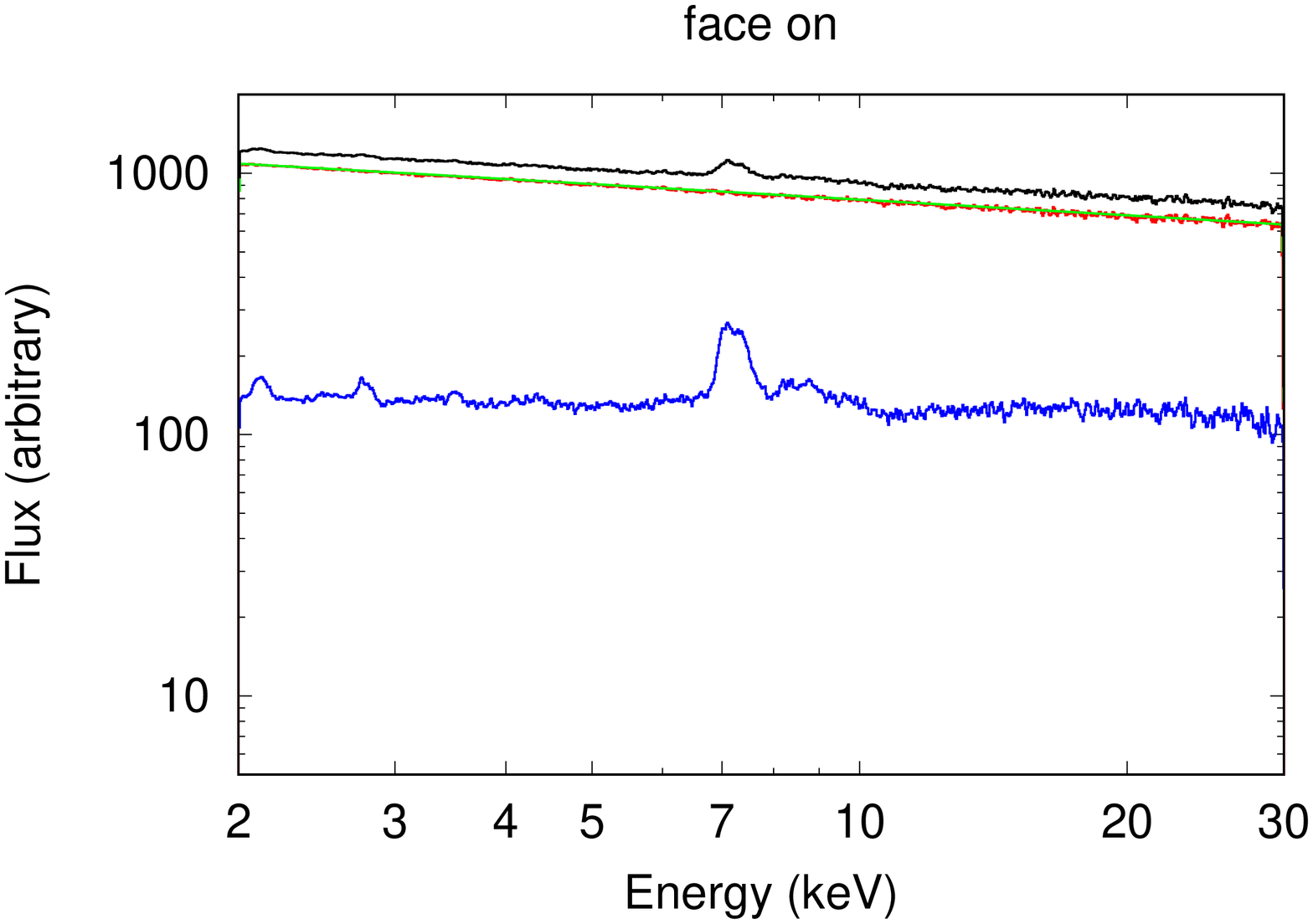}}
\resizebox{5.7cm}{!}{\includegraphics[width=3.2in,angle=270]{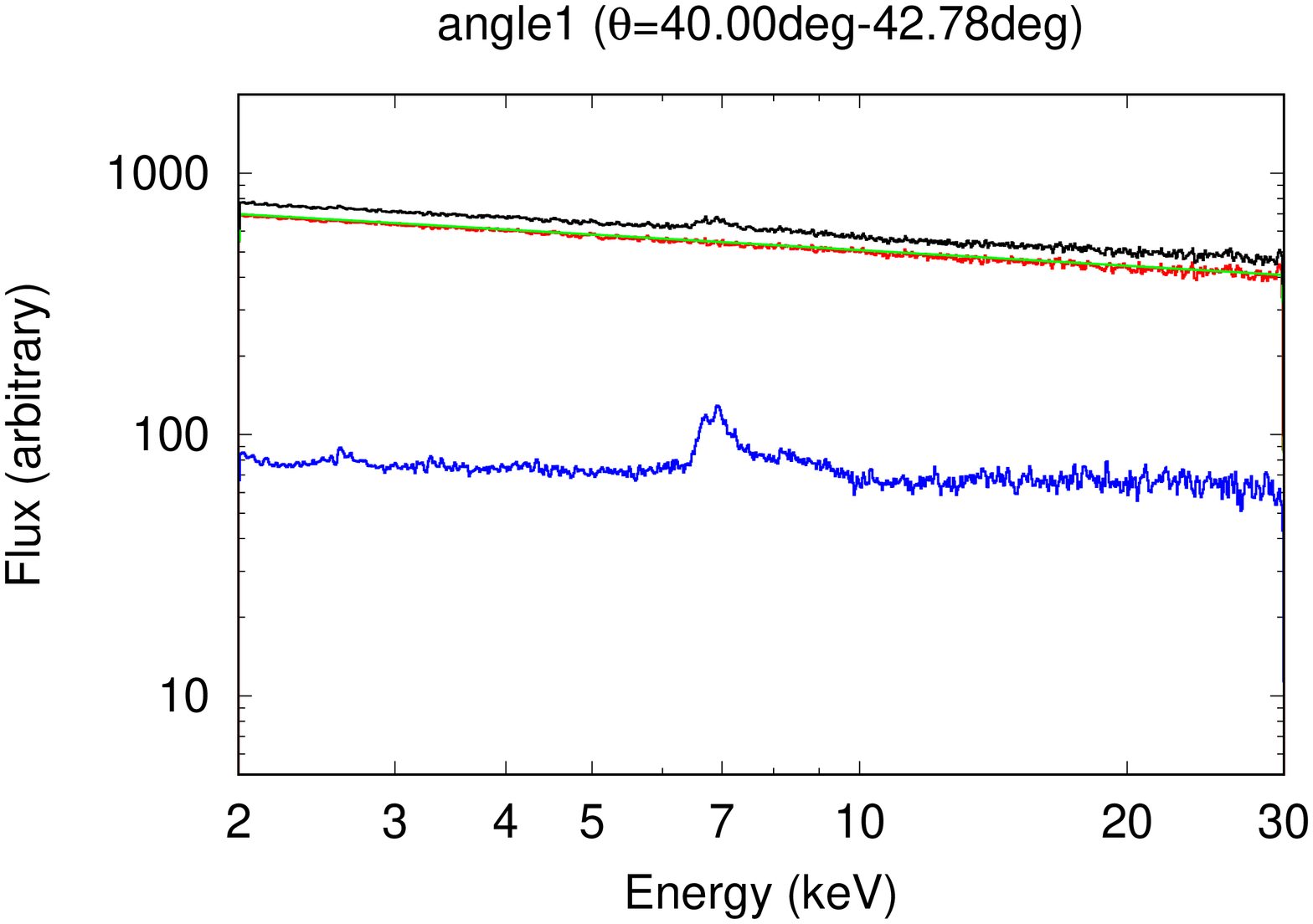}}
\resizebox{5.7cm}{!}{\includegraphics[width=3.2in,angle=270]{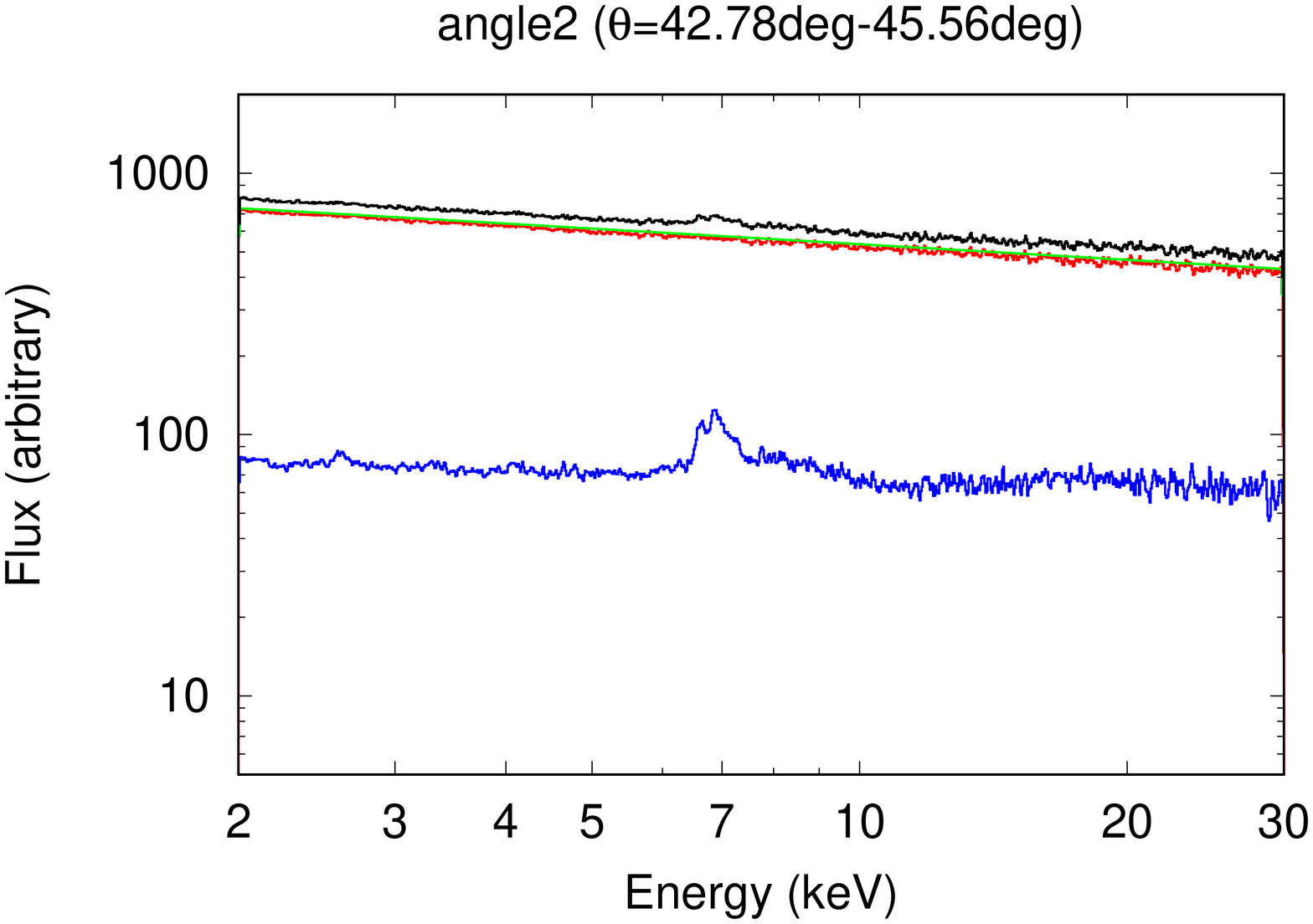}}
}
\subfigure{
\resizebox{5.7cm}{!}{\includegraphics[width=3.2in,angle=270]{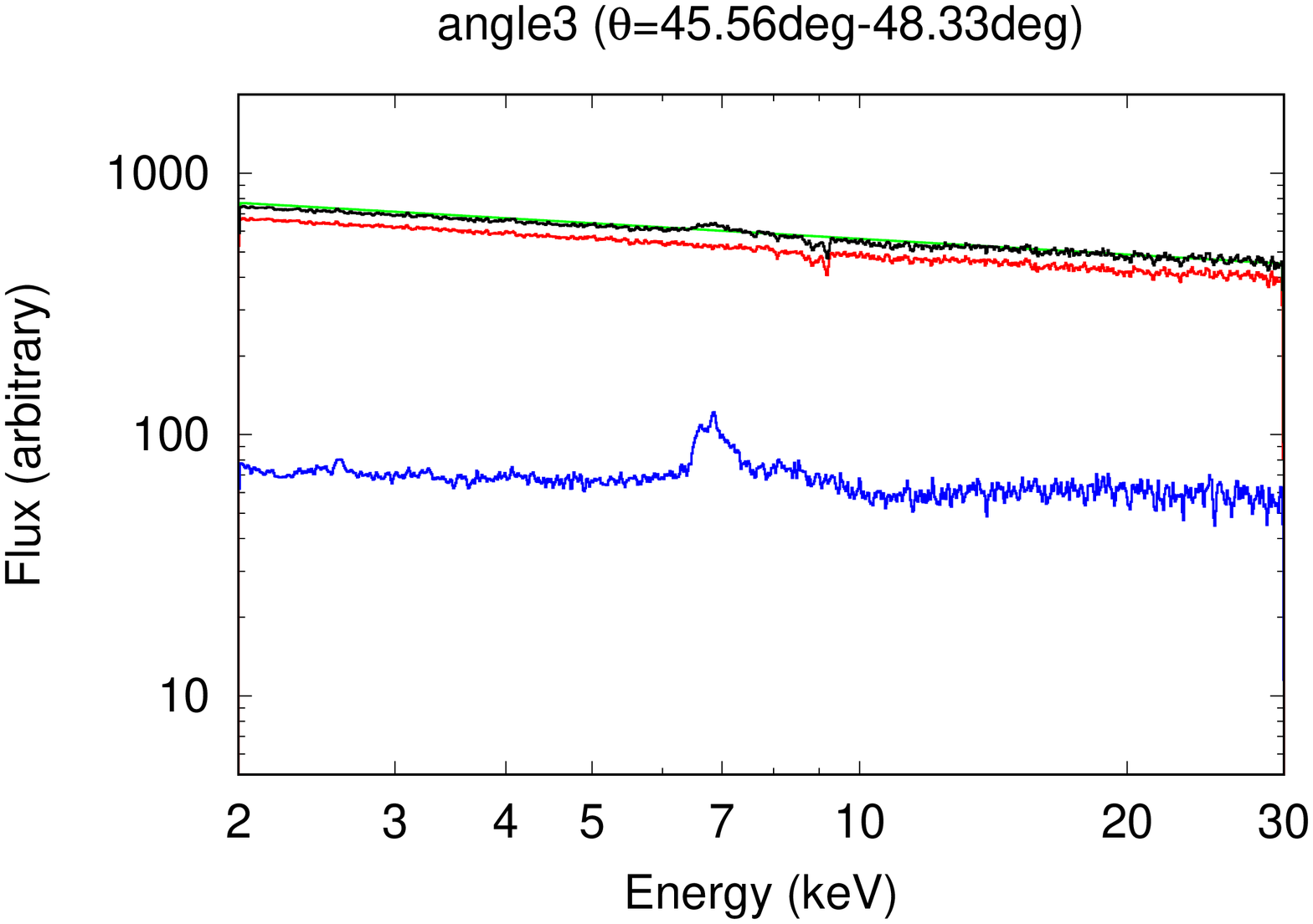}}
\resizebox{5.7cm}{!}{\includegraphics[width=3.2in,angle=270]{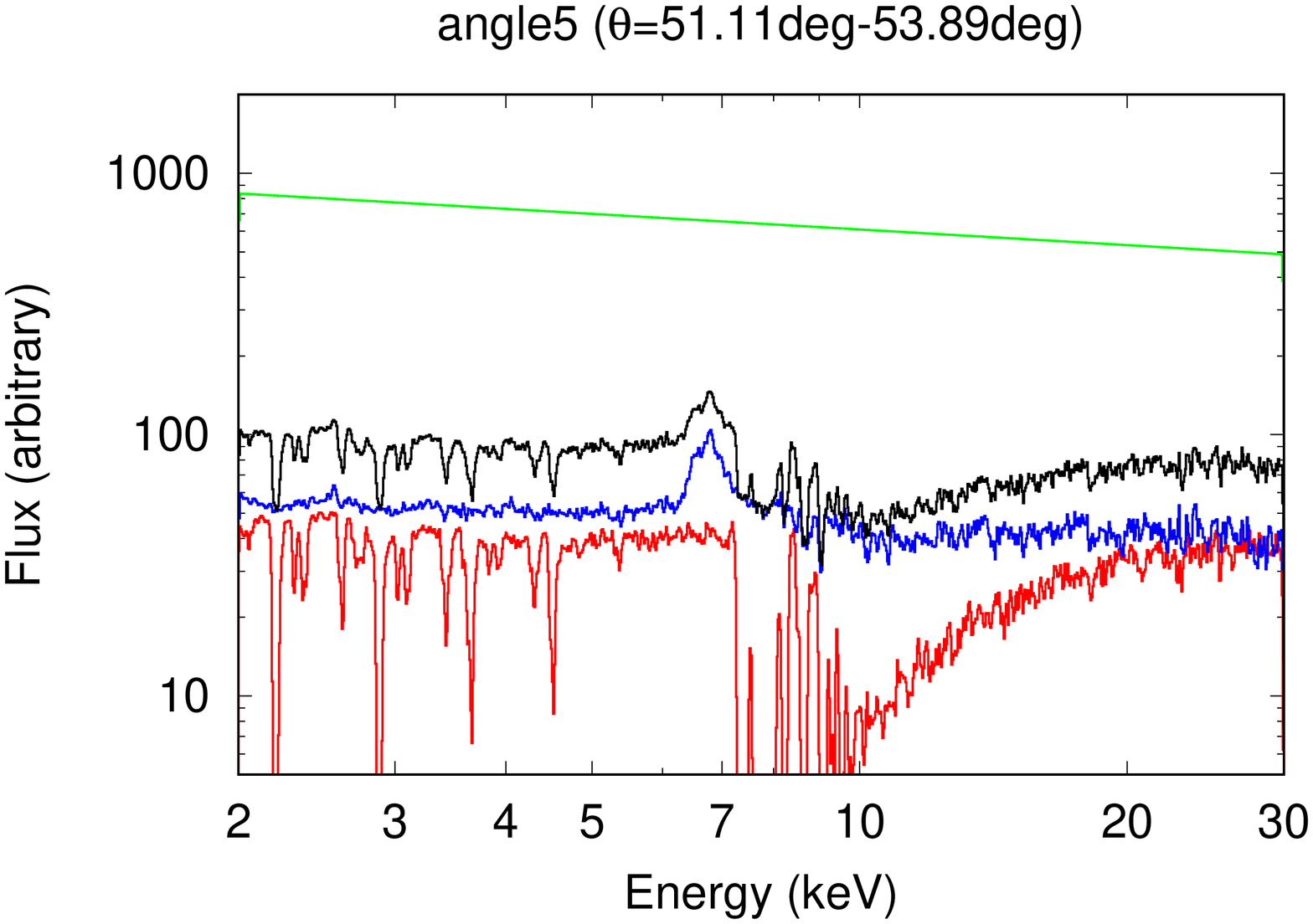}}
\resizebox{5.7cm}{!}{\includegraphics[width=3.2in,angle=270]{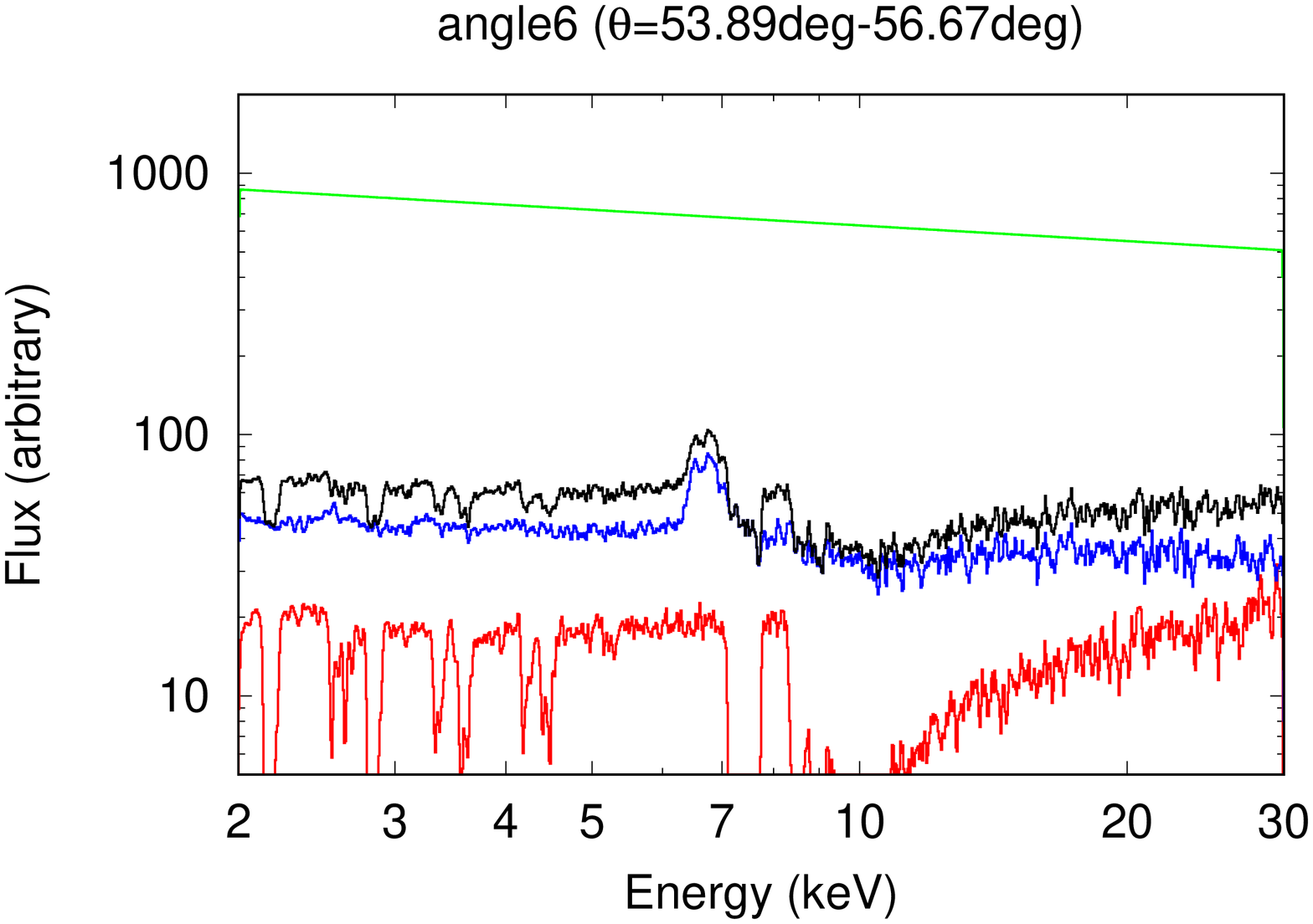}}
}
\subfigure{
\resizebox{5.7cm}{!}{\includegraphics[width=3.2in,angle=270]{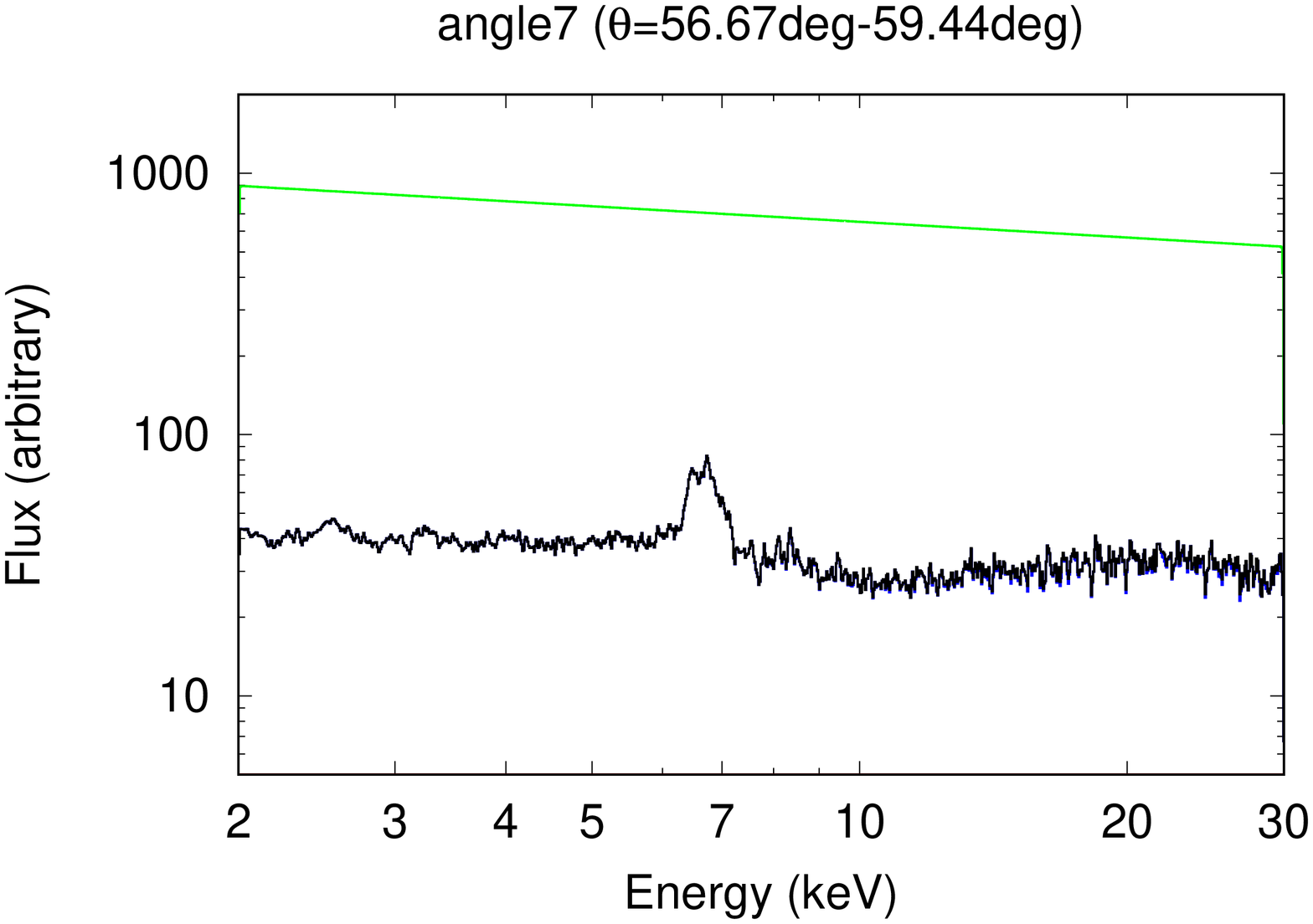}}
\resizebox{5.7cm}{!}{\includegraphics[width=3.2in,angle=270]{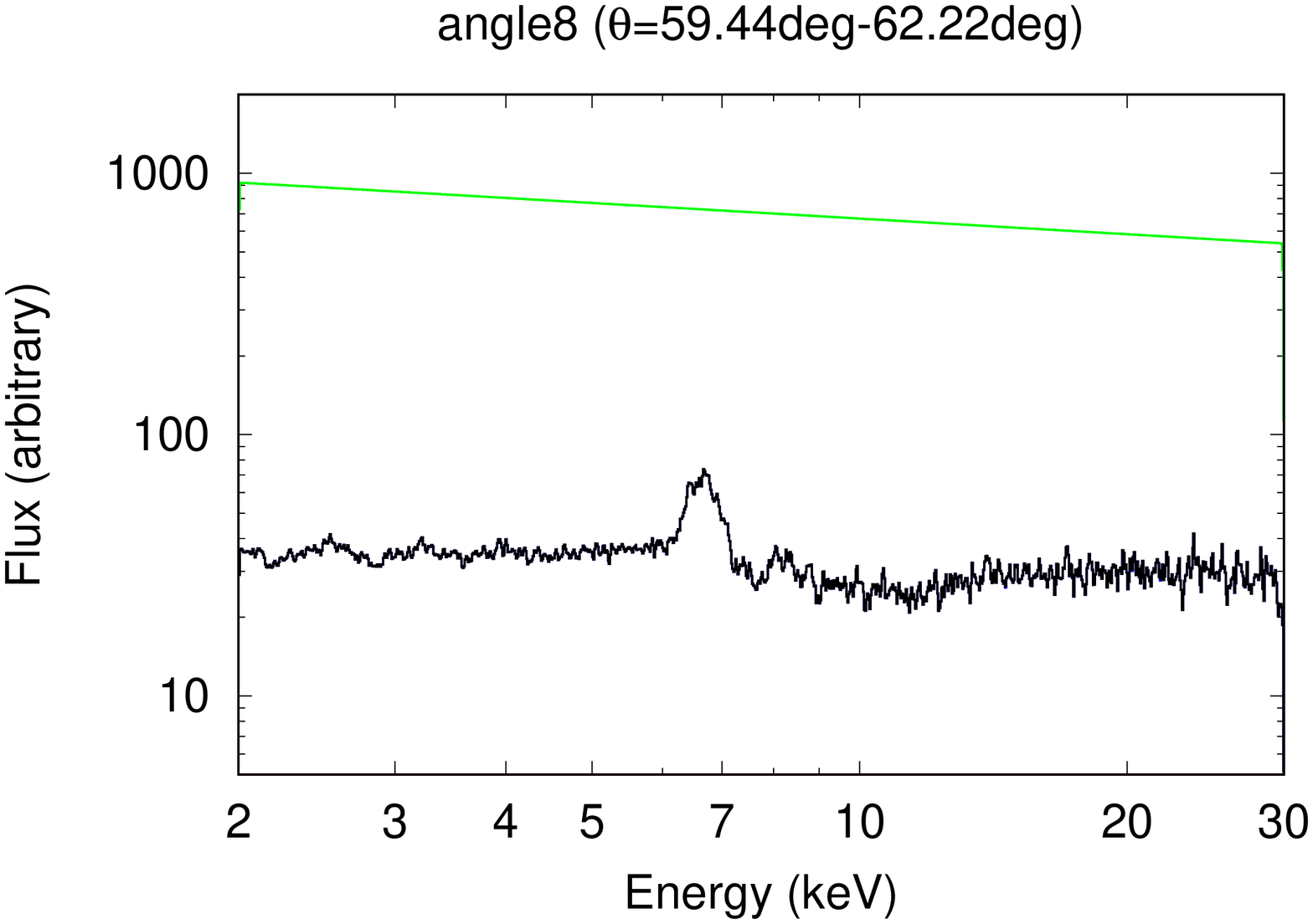}}
\resizebox{5.7cm}{!}{\includegraphics[width=3.2in,angle=270]{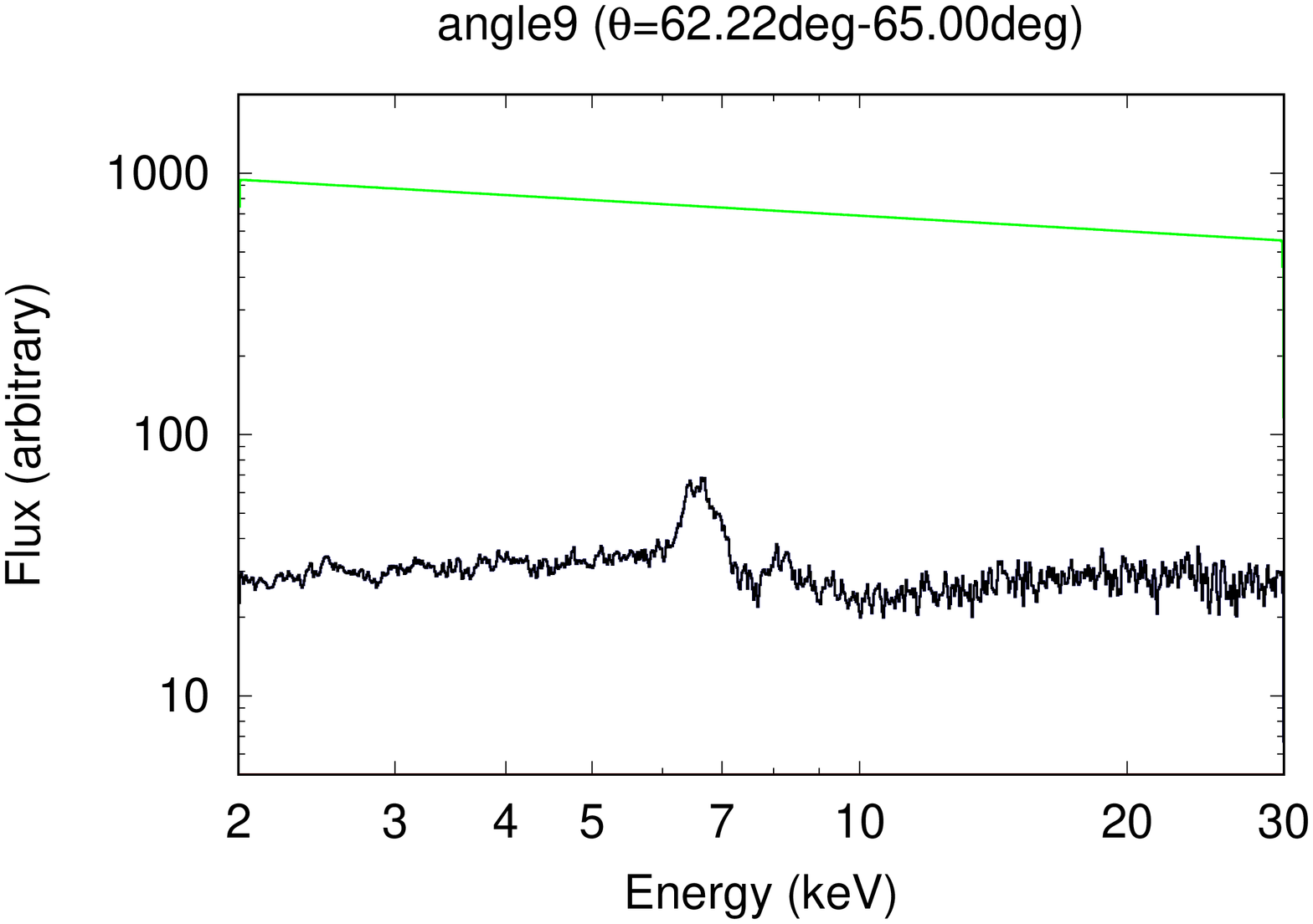}}
}
\caption{
Synthesised X-ray spectra in all the viewing angle case, except for angle4 (see Figure \ref{fig:spectrum_theta4})}
\label{fig:allspectra}
\end{center}
\end{figure*}

The resultant spectra from all other viewing angles are shown in Figure \ref{fig:allspectra}.
The low inclination spectra (up to angle3) have no/little absorption features because the line of sight does not intersect the predominantly equatorial disc wind structure.  Hence the total spectrum is dominated by the primary continuum, though there is also a small contribution from the scattered continuum, which is fainter than the transmitted flux by about a factor 10.
In angle4 (shown previously over this energy range 
in Figure \ref{fig:spectrum_theta4}), only the highly ionised and fast wind exists in the line of sight. The column in the wind is high enough to substantially suppress the transmitted continuum so that the scattered continuum becomes comparable to the primary one.
In angle5 and 6, the highly ionised gas becomes optically thicker and electron scattering reduces the transmitted flux by an even larger factor. This results in a total continuum which is dominated by the scattered spectrum, so the absorption lines and edges are heavily diluted. At even higher inclination angles the line of sight penetrate the X-ray shielding gas and 
the primary spectra are heavily absorbed at soft X-ray energies as well as attenuated by electron scattering. Thus
the total spectra are almost identical to the scattered spectra.

\begin{figure}
\centering
\includegraphics[width=2.2in,angle=270]{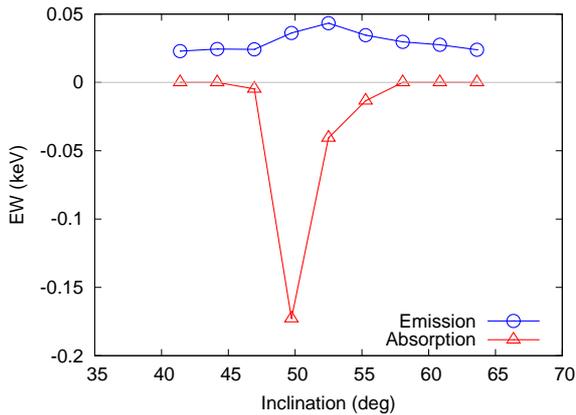}
\caption{Equivalent width for the Fe-K emission/absorption features, calculated in 6.4--7.2/7.4--9~keV, respectively.}
\label{fig:EW}
\end{figure}

Figure \ref{fig:EW} shows EW of the Fe-K emission/absorption features. We fit the continuum (2--5~keV and 20--30~keV) for each viewing angle 
with a power law, then numerically integrate the difference between the model and the simulated data. 
The EW of the emission line is calculated
by integrating over 6.4--7.2~keV, while the summed absorption is integrated from 7.4--9~keV.
The emission EW becomes larger when the electron scattering is stronger, that is, when the primary continuum is less.
The absorption EW is strongest in angle4, where the UFO features are most clearly seen.

\section{Application to PG 1211+143 {\it XMM-Newton} observation}\label{sec4}

We apply our disc wind model to the observational data.
We choose PG 1211+143 ($z=0.081$) as a `canonical' 
powerful UFO target \citep{pou03,ree03,ree09}.
We analyse the European Photon Imaging Camera (EPIC)-pn \citep{epic} archival data (ID: 0112610101, 49.0~ks exposure time) in the {\it XMM-Newton} satellite \citep{xmm}.
The data are reduced using the {\it XMM-Newton} Software Analysis System ({\sc sas}, v.17.0.0) and the latest calibration files. 
In the EPIC analysis, high background periods are removed and 
the source spectrum is extracted from a circular region of $r=30^{\prime\prime}$,
whereas the background spectra are from a circular region $r=45^{\prime\prime}$ in the same CCD chip near the source region without chip edges nor serendipitous sources. The data are 
binned to have a minimum of 40 counts per energy bin.
We use {\tt om2pha} to extract OM spectra.
The spectrum is fitted using {\sc xspec} v.12.10.1 \citep{xspec},

\subsection{UV--X-ray SED}

\begin{figure}
\centering
\includegraphics[width=3in,angle=270]{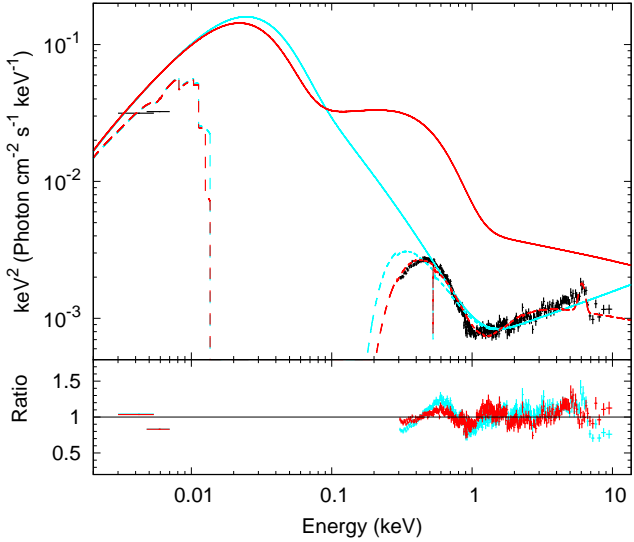}
\caption{
The broad band SED of PG 1211+143 (ObsID=0112610101).
The cyan lines show the naive fitting with the {\tt agnslim} model, with $\Gamma_{\rm hot}=1.6$. The solid line shows the intrinsic SED, whereas the dashed one is with the Galactic absorption.
In order to set $\Gamma_{\rm hot}$ in a reasonable range for a high Eddington object, we introduce partial covering of neutral material, which is shown in red. We also take electron scattering due to the ionised wind into consideration.
The lower panel shows ratios of the data to the model.
}
\label{fig:sed}
\end{figure}

First, we study the broad band SED of this target
to study its mass accretion rate and UV power, and to judge whether this target is suitable for our study.
We use {\tt agnsed} for the AGN SED model \citep{kub18}.
We assume a black hole mass of $\log(M_{\rm BH})=7.83$ (\citealt{bac09}, but not strictly constrained) and co-moving distance of $D=342$~Mpc for redshift of $0.081$
\citep{mar96}.
The EPIC-pn data is binned to have a minimum of 200 counts per energy bin only for this fitting.

We first perform a naive fitting under an assumption that the intrinsic spectrum is absorbed and reddened only by the Galactic column in this direction 
of $N_{\rm H}=2.7\times 10^{20}$~cm$^{-2}$ (HI4PI map; \citealt{HI4PI}), which is modelled by {\tt TBabs} \citep{tbabs}, corresponding to 
$ E(B-V)=0.05$ using 
{\tt redden} \citep{car89} with $E(B-V)=1.7\times N_{\rm H}/10^{22}$~cm$^{-2}$.
We use the abundance table in \citet{tbabs}.
The host galaxy contamination is negligible for such a bright quasar
(see also \citealt{ben09}).
The cyan line in Fig. \ref{fig:sed} shows the fitting result to the broadband spectrum, with almost Eddington mass accretion rate ($\dot{M}_{\rm BH}=0.94\dot{M}_{\rm Edd}$).
The other powerful UFO source in the local universe, PDS 456, has an optical/UV spectrum which similarly implies that the source is around Eddington \citep{hag15,mat16}.
It seems evident that these sources can power winds from continuum radiation pressure alone, but the 
inferred broadband spectral peak in the UV means that UV line driving is also 
expected in these objects, producing even stronger winds than continuum driving alone. 

The model fit to the broadband SED has a soft X-ray excess, fit by 
warm Comptonisation with $kT_{\rm e}=0.20$~keV
and $\Gamma_{\rm warm}=3.2$ (steeper than expected for a purely passive disc, 
but there is plainly strong intrinsic UV disc emission).
The normalisation of the warm Comptonisation component requires all the gravitational power dissipated between $R_{\rm hot}=7.4\,R_g$ and $R_{\rm warm}=2R_{\rm hot}$ (see \citealt{kub18}). 
Below $R_{\rm hot}$ the energy is instead dissipated in a hot Comptonised component, with $\Gamma_{\rm hot}=1.6$.
This is a very hard spectral index for a high Eddington fraction AGN (see e.g.\ \citealt{she06}), and there are clear residuals at Fe K. 
Similarly hard spectra and strong Fe K$\alpha$ features are seen in the `complex' 
NLS1 (\citealt{gal06}, see also \citealt{hag15,hag16}). Currently, these are alternatively modelled either by strong gravitational effects giving the apparently hard spectra through dominance of reflection (e.g.~\citealt{fab09}), or complex absorption (e.g.\ \citealt{mil07,tur07}), but in both cases the intrinsic X-ray emission is steeper and brighter than that observed.
Since we know that there is a wind in PG 1211+143, it seems most likely that there is indeed complex absorption in the line of sight. The strongest UFO wind system known, PDS 456, shows clearly that its 2-10~keV X-ray spectrum requires both highly ionised and more neutral absorption.

\S\ref{sec3} showed that the continuum flux is reduced by electron scattering due to the highly ionised disc wind gas with $N_{\rm H}=1.2\times10^{24}$~cm$^{-2}$ ($r=600-3000\,R_g$ in Figure \ref{fig:tau_a4}).
We fix the column at this value, and model its effect with {\tt cabs}. 
More neutral absorption is often approximated by partial covering \citep{ree03}. 
Therefore, we introduce {\tt zpcfabs} with fixed $\Gamma_{\rm hot}=2.2$ as is typical of AGN with $L\sim L_{\rm Edd}$. We also introduce a Gaussian to fit the Fe-K emission line, with the fixed line width ($\sigma=0.3$~keV).
The resulting parameters of {\tt zpcfabs} are $N_{\rm H}=2.8\times10^{22}$~cm$^{-2}$ with the covering fraction of 0.66.
The overall mass accretion rate is almost unchanged ($\dot{M}_{\rm BH}=0.95\dot{M}_{\rm Edd}$), but the
brighter emission requires more power in the hot Comptonisation, with $R_{\rm hot}=8.4R_g$, and a different shape of the warm Comptonisation, with $\Gamma_{\rm warm}=2.0$.
In both cases, the total mass accretion rate is unchanged since it is dominated by the UV flux. Thus we confirme that PG 1211+143 has similar BH mass and mass accretion rate as those used in the hydrodynamical simulation ($M_{\rm BH}=10^8M_\odot$, $\dot{M}_{\rm BH}=0.9\dot{M}_{\rm Edd}$), and its 
SED is also consistent with the one used in the simulation (Figure 6 in \citealt{nom20}), assuming partial covering (red line).

\subsection{EPIC-pn spectral fitting}
\label{sec4.2}

\begin{figure}
\centering
\includegraphics[width=3.3in,angle=270]{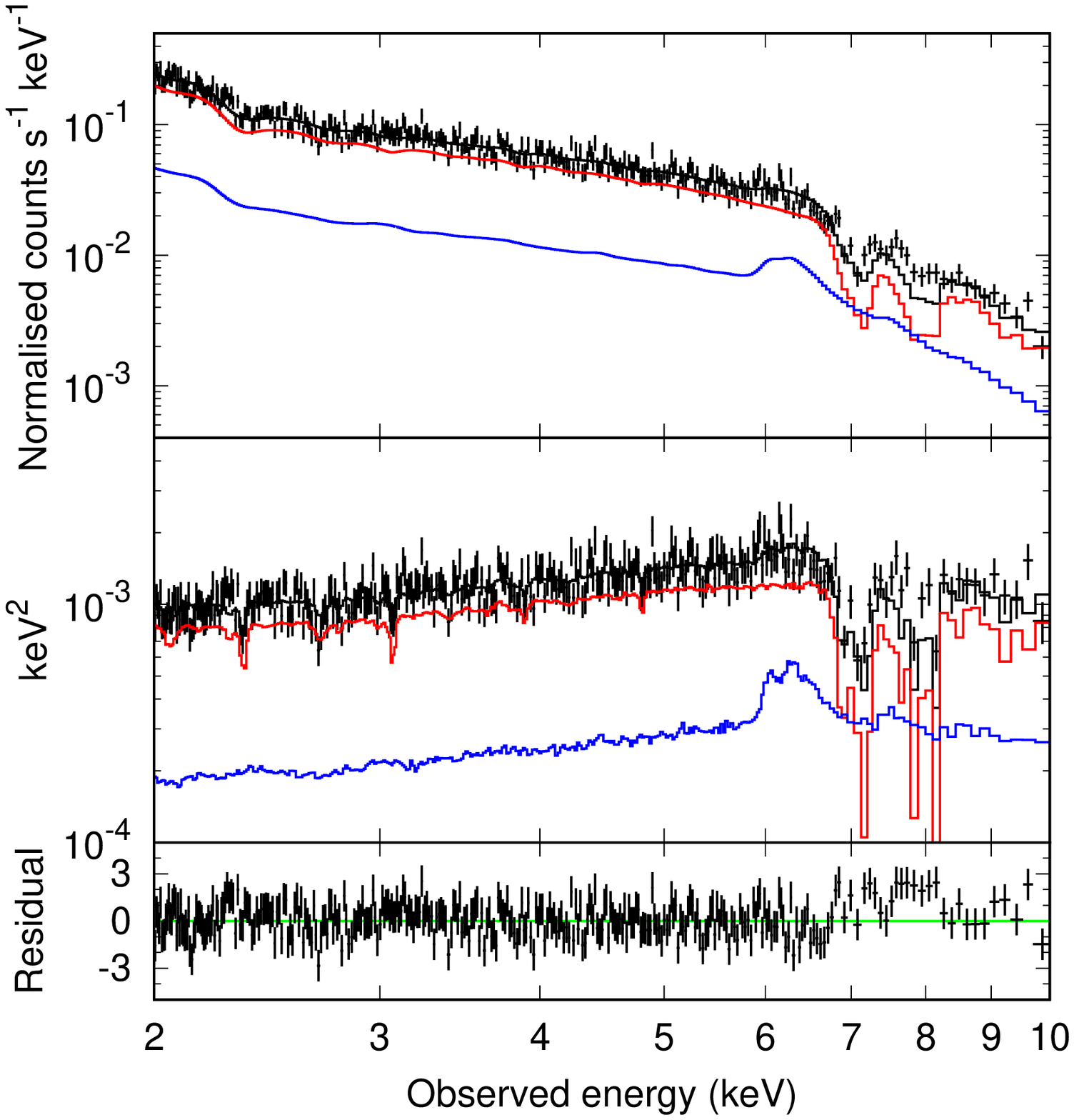}
\includegraphics[width=3.3in,angle=270]{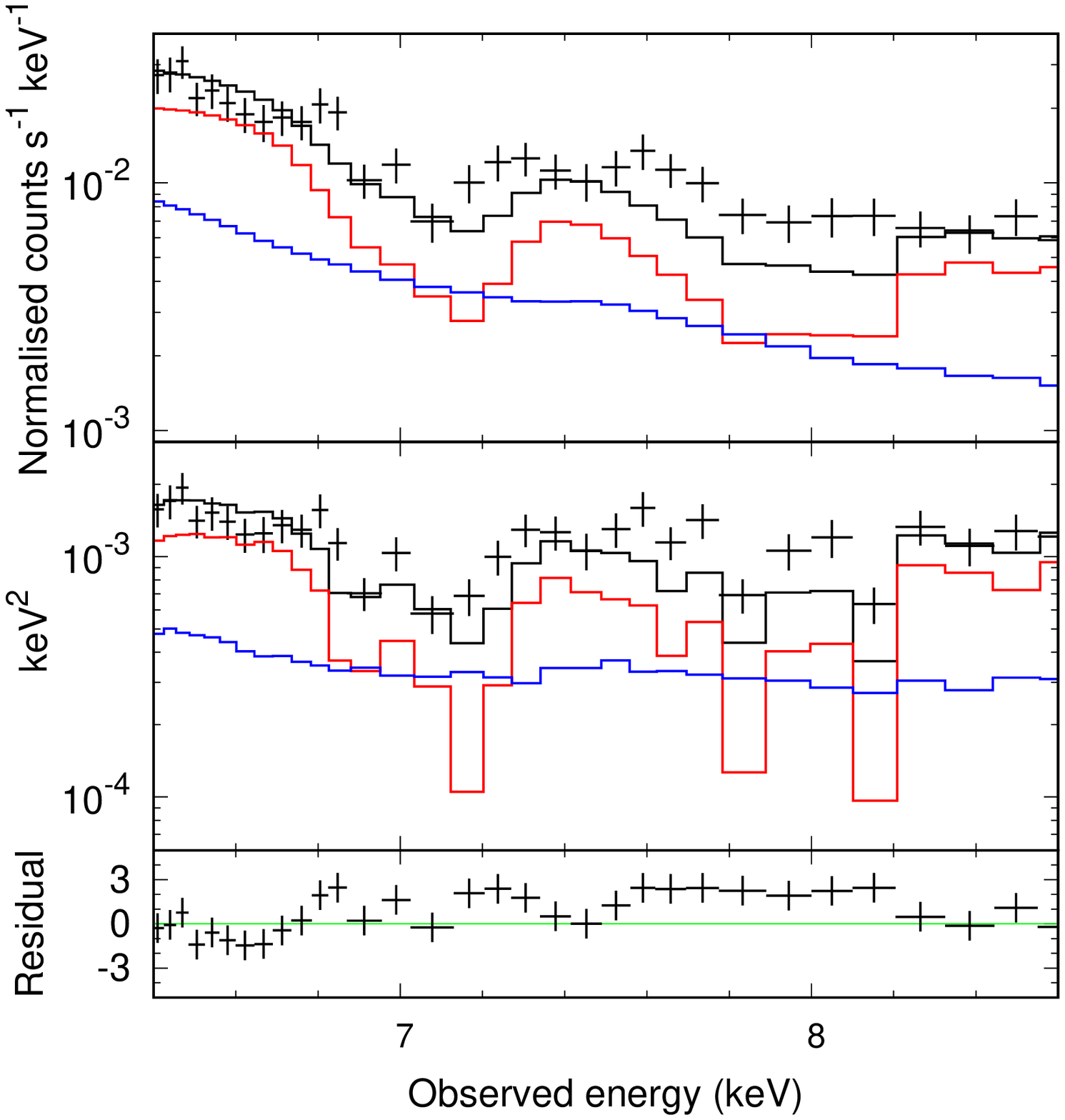}
\caption{X-ray spectrum of PG 1211+143 (ID=0112610101), fitted with our disc wind model.
The upper, middle, and lower panels show the normalised counts (s$^{-1}$ keV$^{-1}$), the keV$^2$ flux (Photons cm$^{-2}$ s$^{-1}$ keV$^{-1}$), and (data$-$model)/error, respectively.
The lower figure is the same as the upper one, but only for 6.4--8.6 keV.
The black bins show the data. The black, red, and blue lines show the total model, primary spectrum, and scattered spectrum, respectively.}
\label{fig:xmm_fitting}
\end{figure}

We fit the EPIC-pn data with the simulated spectrum from angle4. 
We use only the 2--10~keV energy range as 
the current {\sc monaco} version only includes absorption/emission from high ionisation species. 
We see a slight difference in the Doppler shift of the absorption line between the data and the model, so we allow the wind velocity to be a free parameter by multiplying with {\tt zashift} (see also \citealt {hag15}). 
The fitting model is ``{\tt TBabs$\times$zpcfabs}\{{\tt zashift}$\times$(primary component)+(scattered component)\}'', where the {\tt zpcfabs} approximates complex absorption from cold clumps in the wind and is mainly responsible for the apparently hard 2--6~keV spectral shape. 
We find these clumps can be described by $N_{\rm H}=(2.3_{-0.4}^{+0.5})\times10^{23}$~cm$^{-2}$ 
with covering fraction of $0.50\pm0.02$, and the overall fit is good  with $\chi^2/\nu=364.57/337=1.08$.
The other viewing angles over- or underestimate the absorption features and thus cannot reproduce the observed spectrum.

The required shift in velocity is $\Delta v/c=0.026\pm0.010$, 
so the wind in the data is around 10\% slower than that in specific wind simulation used here. Nonetheless, the depth of the strongest absorption feature around 7~keV is well matched by that predicted by the slower wind, but a closer inspection of the residuals reveals that the faster wind overpredicts the depth of the 8~keV absorption lines, though their line energies are well described. 
This is not unexpected as our simulation is only one realisation of a time dependent structure, and the detailed properties of the wind depend somewhat on the assumptions
of radiation transfer (see e.g.\ \citealt{que20}).

As well as matching the absorption features, the 
model also quantitatively explains the broad emission line in the data. The observed emission line 
is consistent with that predicted by scattering 
in the wind material out of the line of sight though there is some room for a small
additional contribution from either a stronger wind or reflection from the inner disc. The red line in Figure \ref{fig:sed} shows the intrinsic SED in this model, without the wind and cold clumps.

\begin{figure}
\centering
\includegraphics[width=2.5in,angle=270]{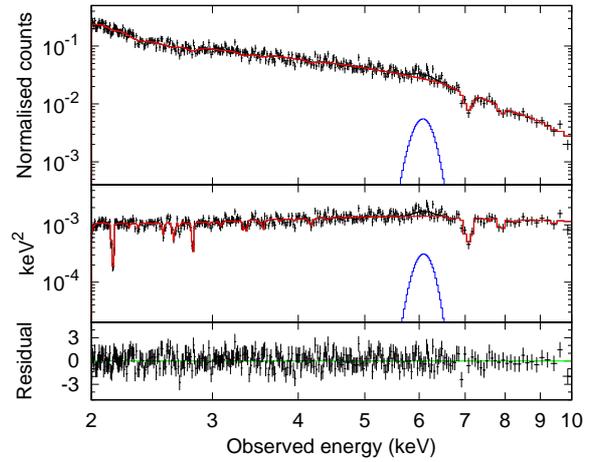}
\caption{X-ray spectrum of PG 1211+143 (same as Figure \ref{fig:xmm3_fitting}, fitted with the photoionised table model
}
\label{fig:xmm3_fitting}
\end{figure}

For the sake of comparison, we fit the UFO in the data with the phenomenological 
photoionised absorber table model, together with a positive Gaussian (Figure \ref{fig:xmm3_fitting}).
We introduced two UFO absorbers, one with $N_{\rm H}=1.2_{-0.3}^{+1.0}\times10^{23}$~cm$^{-2}$, $\log\xi=3.20_{-0.14}^{+0.24}$, $v=(-0.136\pm0.006)c$, and the other with $N_{\rm H}=8.8_{-8.1}^{+17.4}\times10^{23}$~cm$^{-2}$, $\log\xi=5.0_{-0.8}$, $v=(-0.183\pm0.010)c$.
In this phenomenological fitting, each absorber is determined by only the single absorption line feature, so there is large systematic errors on $\log\xi$ depending on whether this line is identified as H or He-like. This gives correspondingly large uncertainties on 
column density and wind velocity.
The $\chi^2/\nu$ is $279.73/341=0.85$, so slightly better, but 
we stress rather that our representative simulation of a UV line driven disc wind from a UV bright, X-ray weak AGN gives rise to a wind which has similar spectral features as seen in the real data from this AGN. 
The snapshot we simulate has a 
large column of highly ionised material with high velocity along the line of sight, which matches quite well to the overall velocity and ionisation structure seen in the data from this `canonical' high power UFO. Plainly UV line driving is capable of producing the observed UFO high ionisation features, despite the low ionisation material required for the acceleration. This can be closer to the disc, out of the line of sight, as suggested by \citet{hag15}.

We conclude that the X-ray spectrum of PG 1211+143 can be fairly well described by the wind model, showing that both the two sets of absorption lines and the strong emission line can be explained as arising in a single UV line driven disc wind structure. 

\section{Simulations of the UV line driven disc wind in the microcalorimeter data}\label{sec5}

Here we simulate observations of the UV line driven disc wind
in various instruments. 
We use our simulation results for angle4 with the PG 1211+143 X-ray flux, without target redshift and additional energy shift.
We simulate X-ray spectra with 100~ks exposures for the current CCD mission, {\it XMM-Newton}/EPIC-pn, and the future mission microcalorimeter data, {\it XRISM}/Resolve and {\it Athena}/X-ray Integral Field Unit (X-IFU). 

\begin{figure}
\centering
\includegraphics[width=3.5in,angle=270]{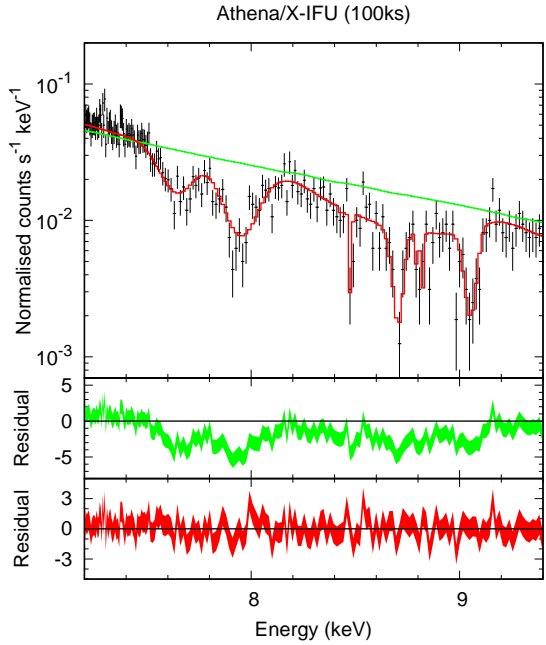}
\caption{
Simulated spectrum for 100~ks {\it Athena}/X-IFU observation.
The green line show the power law continuum, and 
the red line show the best fit of the photoionised absorber model.
The middle and low panels show residuals for the power law model and the best fit photoionised absorber model, respectively.
}
\label{fig:athena}
\end{figure}

\begin{figure}
\centering
\includegraphics[width=3.5in,angle=270]{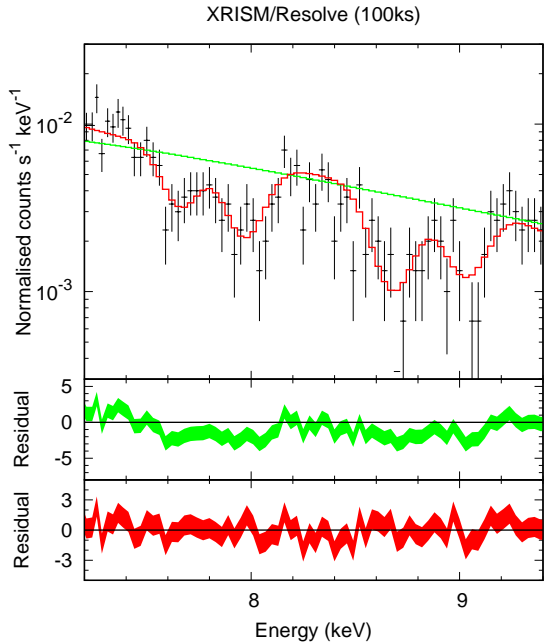}
\caption{Same as Figure \ref{fig:athena}, but for  {\it XRISM}/Resolve
}
\label{fig:xrism}
\end{figure}

\begin{figure}
\centering
\includegraphics[width=3.5in,angle=270]{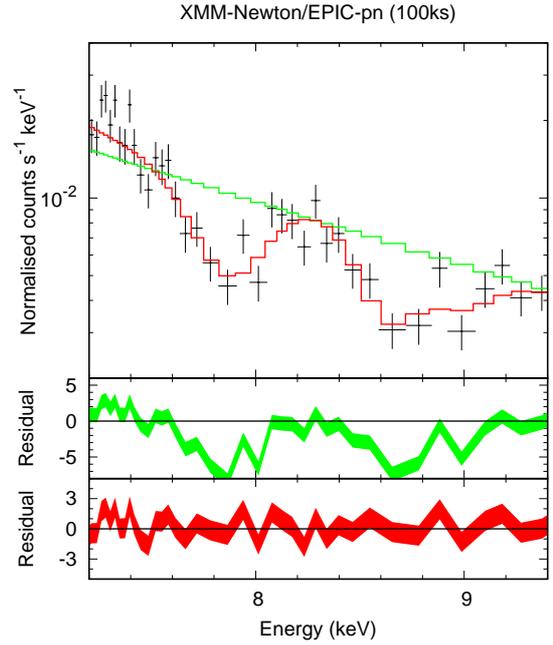}
\caption{Same as Figure \ref{fig:athena}, but for {\it XMM-Newton}/EPIC-pn
}
\label{fig:xmmfak}
\end{figure}

\subsection{Spectral fitting}

First we inspect the {\it Athena} simulation, as this has the highest spectral resolution and photon statistics among the three.
Figure \ref{fig:athena} shows the simulated spectrum and fitting models.
The fitting is performed using the no-binned data based on C statistic \citep{cash} in 5--10 keV, whereas the shown spectrum is binned to have a minimum of 40 counts per energy bin in 7.0--9.4~keV for the sake of visibility.
Complex absorption line features are clearly resolved as residuals of the power law continuum (green line).
We use the {\tt kabs} model for the fitting and estimate $N_{\rm H}$ and $\log\xi$ using \citet{kri18}, which is the same manner as Table \ref{tab:smooth}.
We need four absorbers to get a good fit,
which correspond to Abs1+2, 3, 4, and 5 in Figure \ref{fig:smooth} and Table~\ref{tab:smooth}.
The fitting results are listed in Table\ \ref{tab:fitting}.
The spiky feature due to Abs4 may not be the real one (see \S\ref{sec3}), and we confirmed that even if we ignore it the fitting parameters of the other absorbers change little (within the error bars).
This simulation shows that Athena can trace almost all of the absorbers due to its energy resolution and effective area within a reasonable exposure time.

{\it XRISM}/Resolve has similarly good energy resolution as {\it Athena}/X-IFU, but has smaller effective area.
Therefore a 100~ks observation on {\it XRISM} cannot fully trace all the absorption features expected (Figure \ref{fig:xrism}).
The spectrum is binned after fitting, similarly to the {\it Athena} case, to have a minimum of 60 counts per energy bin.
The absorption features at 8.4--9.1~keV (due to the faster wind) are merged with together, and the statistics only need two absorbers which correspond to the slower wind and the faster wind.
The H-like and He-like Fe ions are clearly resolved and thus the ionisation parameters can be constrained well, which is vital for revealing nature of the wind.

The energy resolution of {\it XMM-Newton} is over a factor 10 worse than {\it XRISM}. We simulate a 100~ks observation of our UV line driven disc wind structure.
The simulated spectrum is binned to have a minimum of 40 counts per energy bin and fit with $\chi^2$ statistics. 
In this case, the data shows only the two broad negative Gaussians at 7.9~keV and 8.7~keV.
It is very difficult to judge whether each (very broad) negative Gaussian is created by both of the H-like and He-like Fe ions with relatively small $\sigma_{\rm v}$ or by a single Fe ion (H-like or He-like) with large $\sigma_{\rm v}$.
{\sc xspec} fitting with the {\tt kabs} model prefers that the line at 7.9~keV is due to the H-like Fe ion with very large $\sigma_{\rm v}=9,000$~km~s$^{-1}$, whereas the one at 8.7~keV is due to both of the H-like and He-like ions with $\sigma_{\rm v}=6,000$~km~s$^{-1}$. Because the slower wind has no He-like Fe ions, we fix a lower limit to $\log\xi$ of 5, which results in a lower limit to the $N_{\rm H}$ of $4.7\times10^{24}$~cm$^{-2}$.
There are clearly large uncertainties on this reconstruction, showing that better spectral resolution is required to determine the wind properties.

\renewcommand{\arraystretch}{1.2}
\begin{table*}
\centering
\caption{Fitting results (Figure \ref{fig:athena}--\ref{fig:xmmfak})}
\label{tab:fitting}
\begin{threeparttable}
\begin{tabular}{llccc}
\hline
\hline
\multicolumn{2}{c}{Model components}  & Athena & XRISM &XMM\\
\hline
Powerlaw & Index & 
&2.2 (fix)&\\
         & Norm ($\times10^{-3}$) \tnote{*1}  &
$2.45\pm0.2$&$2.47\pm0.08$&$2.3\pm0.5$\\
Gaussian & Energy (keV) \tnote{*2}  	
&$6.66\pm0.05$&$6.77_{-0.14}^{+0.16}$&$6.67\pm0.07$\\
         & $\sigma$  (keV) 	
& &0.3 (fix)  &   \\
         & Norm  ($\times10^{-6}$) \tnote{*3}
& $7.0\pm0.8$& $6.7\pm0.3$  & $9.0\pm1.7$  \\
\hline
Abs1+2 & $N_{\rm atom,\,H-like\,Fe}$ ($10^{18}$~cm$^{-2}$) 
&$5.2\pm0.6$ &$4.9\pm1.2$&$8.0_{-1.6}^{+1.3}$\\
& $N_{\rm atom,\,He-like\,Fe}$ ($10^{18}$~cm$^{-2}$) 
&$1.6\pm0.2$ &$1.9\pm0.5$&$0^{+0.7}$\\
& $v_{\rm sigma}$ (km~s$^{-1}$) 
& $4500\pm400$&$5000_{-800}^{+1100}$&$9000\pm2000$\\
& $v_{\rm w}/c$ 
& $-0.1235\pm0.0011$&$-0.126\pm0.003$&$-0.114\pm0.004$\\
\cline{2-5}
& $N_{\rm H}$ ($10^{23}$~cm$^{-2}$) 
&4.8 &3.6&47\\
& $\log\xi$
&4.4&4.3&5 (fix)\\
\hline
Abs3 & $N_{\rm atom,\,H-like\,Fe}$ ($10^{18}$~cm$^{-2}$) 
&$4.5\pm1.9$ &$3.9_{-1.5}^{+1.8}$&$3_{-3}^{+9}$\\
& $N_{\rm atom,\,He-like\,Fe}$ ($10^{18}$~cm$^{-2}$) 
&$1.6_{-0.4}^{+0.8}$ &$3.1\pm0.8$&$2.7_{-0.8}^{+1.5}$\\
& $v_{\rm sigma}$ (km~s$^{-1}$) 
&$10000\pm3000$ &$4900_{-1400}^{+3500}$&$6000_{-2000}$\\
& $v_{\rm w}/c$ 
&$-0.209\pm0.009$ &$-0.228_{-0.002}^{+0.008}$&$-0.223_{-0.011}^{+0.030}$\\
\cline{2-5}
& $N_{\rm H}$ ($10^{23}$~cm$^{-2}$) 
&3.2 &3.3&2.6\\
& $\log\xi$
&4.3&4.2&4.2\\
\hline
Abs4 & $N_{\rm atom,\,H-like\,Fe}$ ($10^{18}$~cm$^{-2}$) 
&$1.8_{-0.5}^{+0.6}$ &---&---\\
& $N_{\rm atom,\,He-like\,Fe}$ ($10^{18}$~cm$^{-2}$) 
&$1.0\pm0.3$ &---&---\\
& $v_{\rm sigma}$ (km~s$^{-1}$) 
&$1200\pm300$ &---&---\\
& $v_{\rm w}/c$ 
&$-0.2304_{-0.0006}^{+0.0005}$ &---&---\\
\cline{2-5}
& $N_{\rm H}$ ($10^{23}$~cm$^{-2}$) 
&1.5 &---&---\\
& $\log\xi$
&4.2&---&---\\
\hline
(Abs5) & $N_{\rm atom,\,H-like\,Fe}$ ($10^{18}$~cm$^{-2}$) 
& $0.6_{-0.3}^{+0.4}$&---&---\\
& $N_{\rm atom,\,He-like\,Fe}$ ($10^{18}$~cm$^{-2}$) 
& $0.28_{-0.13}^{+0.17}$&---&---\\
& $v_{\rm sigma}$ (km~s$^{-1}$) 
& $250_{-70}^{+120}$&---&---\\
& $v_{\rm w}/c$ 
& $-0.2094\pm0.0003$&---&---\\
\cline{2-5}
& $N_{\rm H}$ ($10^{23}$~cm$^{-2}$) 
&0.5 &---&---\\
& $\log\xi$
&4.3&---&---\\
\hline
\multicolumn{2}{l}{C-stat or $\chi^2$ /degree of freedom} & 12311.0/12480 & 8088.5/9988 & 216.3/191 \\
\hline
\end{tabular}
\begin{tablenotes}
\item[*1] photons~keV$^{-1}$~cm$^{-2}$~s$^{-1}$ at 1 keV
\item[*2] in the rest frame, i.e., after redshift correction
\item[*3] total photons~cm$^{-2}$~s$^{-1}$ in the line
\end{tablenotes}
\end{threeparttable}
\end{table*}
\renewcommand{\arraystretch}{1}

\renewcommand{\arraystretch}{1.2}
\begin{table*}
\centering
\caption{Wind properties derived from the photoionised absorber model 
}
\label{tab:fitting2}
\begin{threeparttable}
\begin{tabular}{llcccc}
\hline
\hline
&  & Athena &XRISM & XMM & Answer\\
\hline
Abs1+2 & $r_{\rm min}/R_{\rm g}$ & 130&120&150&$\sim800$\\
 & $\dot{M}_{\rm w}/\dot{M}_{\rm Edd}$  
 &$1.2\times10^{-1}$ &$0.9\times10^{-1}$&$1.3$&---\\
 & $\dot{P}_{\rm w}/(L_{\rm Edd}/c)$  
 &$2.6\times10^{-1}$ &$1.9\times10^{-1}$&$2.5$&---\\
& $L_{\rm w}/L_{\rm Edd}$ 
& $1.6\times10^{-2}$ &$1.2\times10^{-2}$&$1.4\times10^{-1}$&---\\
Abs3 & $r_{\rm min}/R_{\rm g}$ & 40&40&40&$\sim1300$\\
 & $\dot{M}_{\rm w}/\dot{M}_{\rm Edd}$    
 &$4.9\times10^{-2}$ &$4.6\times10^{-2}$&$3.6\times10^{-2}$&---\\
 & $\dot{P}_{\rm w}/(L_{\rm Edd}/c)$   
 &$1.7\times10^{-1}$ &$1.8\times10^{-1}$&$1.4\times10^{-1}$&---\\
& $L_{\rm w}/L_{\rm Edd}$   
 &$1.8\times10^{-2}$ &$2.0\times10^{-2}$&$1.6\times10^{-2}$&---\\
Abs4 & $r_{\rm min}/R_{\rm g}$ & 40&---&---&$\sim2000$\\
 & $\dot{M}_{\rm w}/\dot{M}_{\rm Edd}$    
 &$2.1\times10^{-2}$ &---&---&---\\
 & $\dot{P}_{\rm w}/(L_{\rm Edd}/c)$  
 &$8.0\times10^{-2}$ &---&---&---\\
& $L_{\rm w}/L_{\rm Edd}$   
 &$0.9\times10^{-2}$ &---&---&---\\
Abs5 & $r_{\rm min}/R_{\rm g}$ &40 &---&---&$\sim2000$\\
 & $\dot{M}_{\rm w}/\dot{M}_{\rm Edd}$    
 &$7.6\times10^{-3}$ &---&---&---\\
 & $\dot{P}_{\rm w}/(L_{\rm Edd}/c)$   
 &$2.7\times10^{-2}$ &---&---&---\\
& $L_{\rm w}/L_{\rm Edd}$   
 &$2.8\times10^{-3}$ &---&---&---\\
\hline
Total  & $\dot{M}_{\rm w}/\dot{M}_{\rm Edd}$  & $2.0\times10^{-1}$&$1.4\times10^{-1}$&$1.3$&$5.6\times10^{-1}$\\
 & $\dot{P}_{\rm w}/(L_{\rm Edd}/c)$  &$5.3\times10^{-1}$  &$3.7\times10^{-1}$& 2.6& 1.7\\
& $L_{\rm w}/L_{\rm Edd}$ &$4.6\times10^{-2}$ & $3.2\times10^{-2}$&$1.6\times10^{-1}$&$1.7\times10^{-1}$ \\
\hline
\end{tabular}
\end{threeparttable}
\end{table*}
\renewcommand{\arraystretch}{1}

\subsection{Physical parameters of the wind}

Next, we try to estimate the mass loss rate, momentum flux, and kinetic power of the wind from the fitting results of the photoionised table model.
The mass loss rate is expressed as
\begin{eqnarray}
\dot{M}_{\rm w}&=&A(r) \rho(r)v(r) \nonumber \\
&=& (\Omega b r^2) (1.2m_{\rm p}n(r))v_{\rm w}, \label{eq:Mdot} 
\end{eqnarray}
where $\Omega$ is the solid angle of the wind ($\sim1.6\pi$), $b$ is the filling factor, $m_{\rm p}$ is the proton mass,  $n(r)$ is the electron number density, and $v_{\rm w}$ is the wind velocity \citep{gof15}.
When the column density is 
\begin{equation}
N_{\rm H}=\int n(r)dr \sim b n(r) r, \label{eq:NH}
\end{equation}
equation (\ref{eq:Mdot}) is calculated as 
\begin{equation}
\dot{M}_{\rm w}\sim\Omega m_{\rm p} v_{\rm w} r N_{\rm H}.\label{eq:Mdot2}
\end{equation}
The momentum flux and kinetic power are $\dot{P}_{\rm w}=\dot{M}_{\rm w}v_{\rm w}$ and $L_{\rm w}=\dot{P}_{\rm w}v_{\rm w}/2$.

The mass loss rate is proportional to the line formation radius of the wind, which is difficult to constrain.
\citet{gof15} discuss estimates of the radius from ionisation state and escape velocity respectively as
\begin{eqnarray}
r_{\rm max} &=& \frac{L_{\rm ion}}{\xi N_{\rm H}}  \label{eq:rmax}\\
r_{\rm min} &=& \frac{2GM_{\rm BH}}{v_{\rm w}^2} \label{eq:rmin}.
\end{eqnarray}

With $L_{\rm ion}=0.6L_{\rm Edd}$, the maximum radius (eq.~\ref{eq:rmax}) becomes implausibly large, $r_{\rm max}>10^4\,R_g$. It is difficult for
any physical mechanism to drive UFOs from such a large radius. Hence the
minimum radius ($r_{\rm min}$, which is same as the wind launching radius) is usually adopted. 
Substituting $r_{\rm min}$ into eq.~(\ref{eq:Mdot2}),
we get the mass loss rate, momentum flux, and kinetic power for each absorber and each spectral fitting, which are shown in Table\ \ref{tab:fitting2}.

The estimated wind properties in the microcalorimeter simulations are smaller than in the
simulation by about a factor 10, where $\dot{M}_{\rm w}/\dot{M}_{\rm Edd}=0.56$,  $\dot{P}_{\rm w}/(L_{\rm Edd}/c)=1.7$, and  $L_{\rm w}/L_{\rm Edd}=0.17$ (see \S\ref{sec2.1}).
Even if we assume that these multi-velocity components are produced by different winds and sum up all the values, the total still dramatically underestimates the true answer.

The main reason of this large discrepancy is underestimation of the 
line formation radius from the data. 
The mass loss rate is proportional to the 
assumed wind launching radius in eq.\ (\ref{eq:Mdot2}).
Observers typically adopt, $r_{\rm min}$, but in our simulation this is an
order of magnitude smaller than the radius at which the line is formed
of $r\sim700-2000\,R_g$ (see Table\ \ref{tab:fitting2}).
This discrepancy becomes larger as the wind velocity becomes faster.
The acceleration from UV line driving means that the fastest wind 
is launched at larger radii in Figure \ref{fig:tau_a4} whereas 
assuming 
a wind launching radius equal to the escape velocity requires smaller $r_{\rm min}$.
This has an especially large impact on the kinetic power estimation.

\begin{figure}
\centering
\includegraphics[width=3.5in,angle=270]{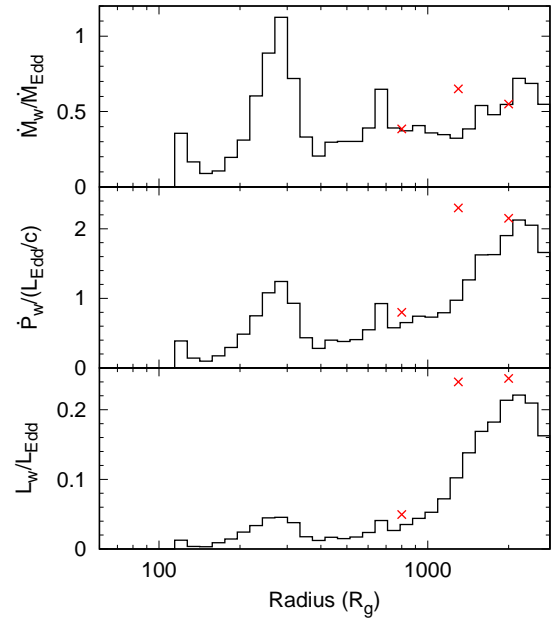}
\caption{mass loss rate, momentum flux, and kinetic power of the wind as a function of radius. The black lines are calculated from the radiation hydrodynamics simulation, whereas the red bins are from the {\it Athena} simulation (Abs1+2, 3, 4), based on the line formation radius (shown in the ``answer'' column in Table \ref{tab:fitting2}) and $\Omega/4\pi=0.2$.}
\label{fig:property}
\end{figure}

We recalculate the wind properties using the line formation radius shown in the ``answer'' column in Table \ref{tab:fitting2}.
In addition, we adopt that $\Omega/4\pi=0.2$ for the radiation-driven wind (see Figure \ref{fig:wind_density}).
We do not consider Abs5 because it spans only a rather small
radius range and thus the assumption in equation (\ref{eq:NH}) is broken.
Figure \ref{fig:property} shows the wind properties as a function of radius.
The corrected properties derived from the {\it Athena} simulation (red crosses) match well to the answer (black lines), within a factor of $\lesssim3$.

In our wind geometry, the single wind makes several components with different velocities.
This means that we should not sum up the wind properties derived from different components, but rather average them to get the typical rates of wind mass loss, momentum and kinetic power. 
This demonstration shows that we can correctly estimate the wind properties using the great resolving power of the microcalorimeter only by combining these data with a wind launching radius based on physical models.

\section{Discussion}\label{sec6}

\subsection{Comparison with previous studies}

The results shown here are the first to explicitly fit the spectral features from the strongest UFO AGN using a UV line driven disc wind simulation. Previous work using a similar post-processing radiation transfer approach on the radiation hydrodynamic simulation of \citet{pro04} showed clearly that UV line driving could produce a high ionisation wind along some lines of sight \citep{sim10b}, but this wind velocity was not really comparable to the UFO velocity, being typically only $\sim 0.05c$
This is not fast enough to match the most powerful UFOs. 
The difference in our study is that we use the density/velocity structure  from the radiation hydrodynamic simulation of \citet{nom20}. The differences in radiation transfer give a wind which is significantly faster, and so is a much better match to the observed properties of the UFO in PG 1211+143. This enables us to directly fit the simulation results to the data, unlike \citet{sim10b} where only the simulation results are shown, though they did apply their radiation transfer code to data using a phenomenological biconical wind structure \citet{sim10a}. 
Thus our paper is the first to directly show that UV line driven wind simulations can fit the observed UFOs.

\subsection{Broad absorption line width}\label{sec6.1}
The UFO absorption lines detected in the CCD detectors ({\it XMM-Newton} and/or {\it Suzaku}) sometimes have broad line widths, up 
to $\sim10,000$~km~s$^{-1}$ (e.g.~PDS 456; \citealt{ree03,ree09}) though 
more typical velocity widths are 
$\sigma_v\lesssim5,000$~km~s$^{-1}$ \citep{tom11}. 
\citet{fuk19} suggested that  this line broadening is due to the transverse velocity gradient in the context of magnetically driven wind.
Magnetically driven winds are strongly connected with disc rotation and have large azimuthal velocity. This introduces line broadening if the 
X-ray corona has radial extent, so that the transverse wind motion across the corona makes a difference to the wind velocity along the line of sight.
They applied their wind model and assumed geometry to the PDS 456 X-ray spectrum and constrained the coronal size as $\lesssim10\,R_g$. They concluded that 
the absorption line broadening is ``a serious challenge to radiation-driven wind viewpoint'' because radiation driven winds have slower azimuthal velocity so cannot make the broad absorption in the same way.

It is true that the absorption lines are not broadened via the small azimuthal velocity of the radiation-driven wind $v\lesssim0.1c$ for almost all of the grids, as pointed out by \citet{fuk19}.
However, radiation driven winds, specifically the UV line driven winds in this simulation, have strong radial velocity gradients, so it is very easy to produce the observed widths of the absorption line from radial rather than transverse motion. We note that our simulation may overestimate the discrete nature of the different absorbers as the finite resolution elements pixellate a 
smoother and more continuous wind acceleration law. However, this effect is countered by the limited resolution of current detectors. For example, 
our simulated {\it XMM-Newton} observation (Figure \ref{fig:xmmfak}) shows that 
several wind components merge together can  be fit with a single Gaussian absorption feature.

Broad absorption lines are a generic feature of the radial structure of UV line driven disc winds. These arise from radial velocity gradients in the wind even for a purely point source of X-rays. They do not require magnetic winds.

\subsection{Impact on AGN feedback}\label{sec6.2}
A robust correlation has been established between SMBH mass and velocity dispersion of the galactic spheroid; it is called as $M-\sigma$ relation (e.g.~\citealt{mag98,sil98}).
The exact mechanism to make this correlation is still to be solved, but
AGN winds such as UFOs are considered to be one of the plausible trigger.
The UFO kinetic power often exceeds the binding energy of the spheroidal component, so
it can quench star formation which is needed to reduce angular momentum of the circumnuclear disc, resulting into suppressing the mass accretion to the accretion disc (e.g.~\citealt{fer10}).
The UFO energy is theoretically considered to be transferred into the galactic-scale cold molecular outflow (e.g.~\citealt{pou13}),
which is observed at the sub-millimetre and far-infrared wavelengths with a size of $\sim$kpc and $v\sim500$~km~s$^{-1}$ (e.g.~\citealt{cic14}).
This type of feedback is called the ``quasar mode'', in contrast to the ``radio mode'', where highly collimated jets take away the kinetic energy.

In order to investigate the detailed mechanism of the quasar mode feedback,
the energy transfer from UFO (sub-pc scale) to molecular outflow (kpc-scale) has been recently studied.
\citet{tom15} compared momentum fluxes of the two outflows in IRAS F11119+3257 and 
showed a physical connection between them for the first time.
Such comparisons have now been performed for several additional sources (e.g.~\citealt{miz19b}).
How the AGN outflow travels have been studied in both the observational and theoretical viewpoint
(e.g.~\citealt{wag13}, \citealt{cic18}, and references therein).
When we quantitatively study how the AGN outflow travels,
calculating the UFO momentum flux, which is calculated by mass loss rate and wind velocity, is critical.
Using the simulation-based model we introduce in this paper,
we can get more reliable UFO parameters,
which is critical for studying the AGN quasar mode feedback process, but only when these are combined with a theoretical understanding of the wind launch radius. 

\section{Conclusion}\label{sec7}
We performed Monte Carlo radiation transfer of a state of the art radiation hydrodynamic UV line driven disc wind to explicitly test whether 
this can explain the observed UFO properties.
UV line driven disc winds had been ruled out as the origin of UFOs 
due to their high ionisation state, where UV opacity is very low. 
Previous radiation hydrodynamic simulations had shown that this was not necessarily true, that there are 
highly ionised sightlines through a UV line driven disc wind because the wind geometry is not spherical \citep{pro04, sim10b}. 
The acceleration region 
is localised near the disc and there are 
lines of sight which only pass through the highly ionised 
at mid-viewing angles ($\theta\sim50$~deg). However, this work was based on one particular radiation hydrodynamic simulation \citep{pro04}, and had a typical velocity which was $\sim 0.05c$, which is much slower than the 'poster child' AGN UFO winds seen in e.g. PG~1211+143 and PDS 456. Here we use instead a newer simulation from \citet{nom20}, where the radiation hydrodynamics includes more realistic X-ray and UV opacity, and iterates to adjust the disc structure and UV luminosity to the reduction in mass accretion rate caused by the wind. The wind in this simulation is must faster, reaching $\sim 0.2c$ as seen in the strongest, most convincing UFOs. We simulate the spectra along different lines of sight using Monte-Carlo post-processing, and show that this 
disc wind includes sight lines where the wind is both highly ionised and fast, as required. The simulation predicts complex absorption features in 7--10~keV in the rest frame for a sight line through the upper part of the wind. 
The primary continuum becomes weaker than the input power law due to electron scattering, so the scattered continuum makes a larger contribution to the total flux. The main effect of this is to dilute  the absorption line depth, but also introduces a broad Fe-K emission line 
due to fluorescence and/or resonance scattering.
We successfully explain the P Cygni like feature in the PG 1211+143 X-ray spectrum with our model. This is the first time that the X-ray data from a UFO has been convincingly fit by direct results from a UV line driven disc wind simulation. 

Our UV line driven disc wind produces absorption lines with multiple velocity components.
Such multi-phase UFOs have been reported in some AGNs other than PG 1211+143, e.g.\ MCG-03-58-007 \citep{bra18} and IRAS 13224--3809 \citep{par20}, and often interpreted by the stratified wind where the faster component is produced from the inner region, and vice versa (see \citealt{par20} and reference therein).
We demonstrate in this paper that the radial velocity structure in the radiation-driven wind naturally results in the multi-phase UFOs.
The wind is accelerated along the line of sight so that the faster component is located farther from the central object.

We also performed simulations of the wind features at microcalorimeter resolution for both XRISM and Athena. Their excellent energy resolution enables us to trace the detailed structure of the wind and reveal the more complex absorption line profiles than we have thought. 
We show that fitting the wind simulation with phenomenological 
photoionised absorbers (as generally used in current studies) 
underestimates the wind properties (mass loss rate, momentum flux, and kinetic power) by one order of magnitude. This is due to the assumption that the wind is launched at a radius where the escape velocity is equal to the observed terminal velocity. However, this wind launching radius ($r\sim100R_g$) is much smaller than the typical line formation radius where the UV line driven disc wind 
makes the absorption lines ($r\sim1000R_g$).
We show that the wind properties can be recovered if the ``correct'' wind radius is used. This shows the importance of using physical wind models to interpret observational data, especially where this is used to input 
AGN feedback power into cosmological simulations.

\section*{Acknowledgements}
This work is supported in part by Japan Society for the Promotion of Science (JSPS) Grant-in-Aid for Scientific Research A (17H01102, KO), for Scientific Research C (16K05309, MN, KO; 18K03710, KO), for Scientific Research on Innovative Areas (18H04592, KO), by Science and Technology Facilities Council (STFC) grant (ST/P000541/1, CD), by Kavli Institute for the Physics and Mathematics of the Universe (IPMU) funding from the National Science Foundation (No.\ NSF PHY17-48958, CD), by the Ministry of Education, Culture, Sports, Science and Technology of Japan (MEXT) as `Priority Issue on Post-K computer' (Elucidation of the Fundamental Laws and Evolution of the Universe, KO), and by Joint Institute for Computational Fundamental Science (JICFuS, KO).
Numerical computations were in part carried out on Cray XC50 at Center for Computational Astrophysics, National Astronomical Observatory of Japan.
MM acknowledges the support from JSPS overseas research fellowship and Hakubi project at Kyoto University.
The authors thank Dr.~Kouichi Hagino and Dr.~Ryota Tomaru for their technical advice on {\sc monaco}, and thank Dr.~Atsushi Tanimoto for making the table model.

\section*{Data Availability}
The {\it XMM-Newton} data and the spectral table model underlying this article are available in the XMM-Newton Science Archive (XSA) and \url{https://github.com/mskmizu/uvwind}, respectively.




\bibliographystyle{mnras}
\bibliography{00} 



\appendix
\section{Snap shots of the disc wind}\label{sec:a1}
\begin{figure}
\centering
\includegraphics[width=2.5in,angle=270]{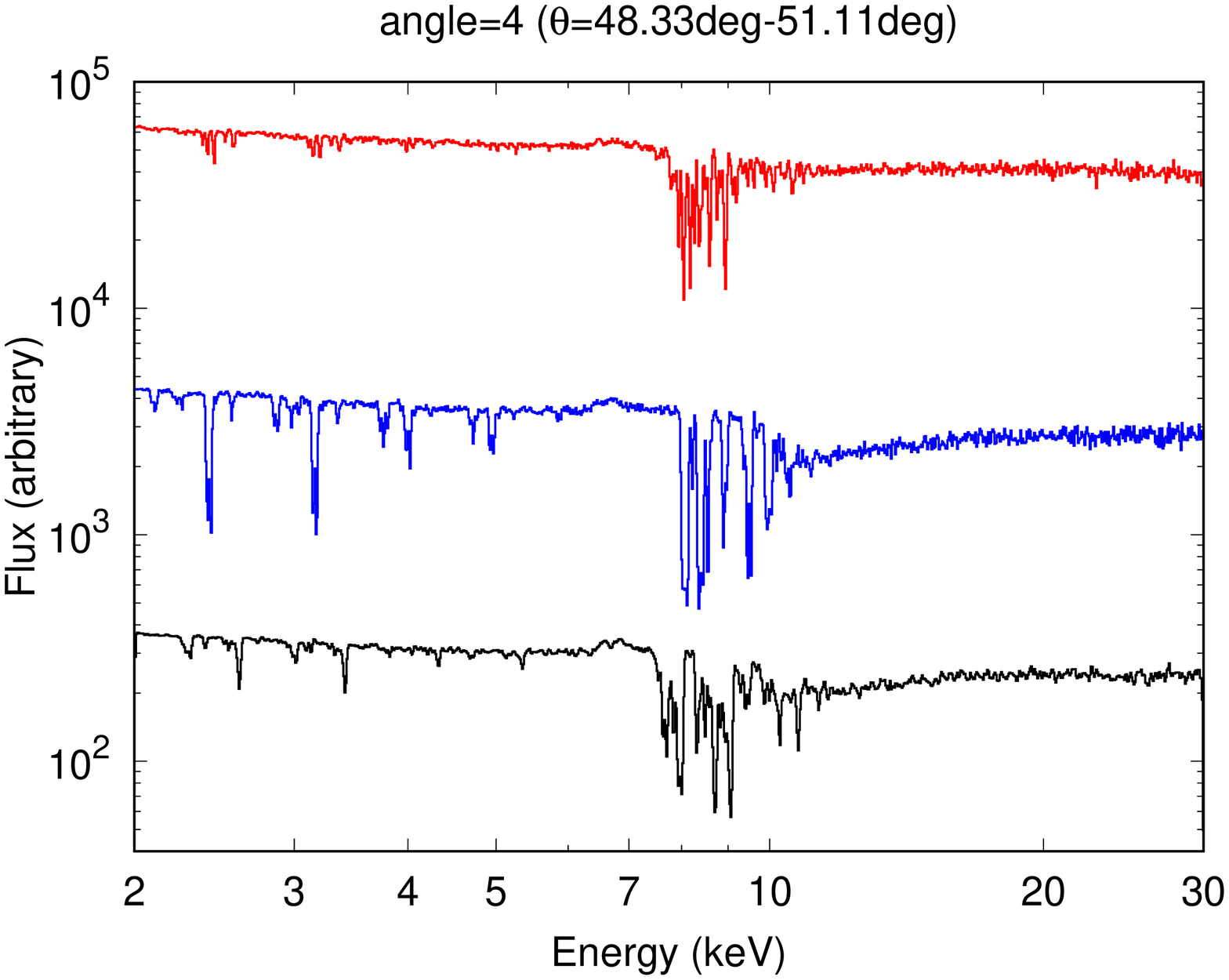}
\caption{Energy spectra with different snap shots, time$=3\times10^5R_g/c$ (red), $4\times10^5R_g/c$ (blue), and $5\times10^5R_g/c$ (black, used in the main text). 
}
\label{fig:compare}
\end{figure}

The disc wind shows time variability.
\citet{nom20} showed the time-averaged disc wind properties to study the parameter dependence, but in this paper
we use snap shots to compare them to the observations.
We constructed three energy spectra for each viewing angle, time after the simulation starts is $3\times10^5R_g/c$, $4\times10^5R_g/c$, and $5\times10^5R_g/c$ (Figure \ref{fig:compare} and Table\ \ref{tab:compare}).
$v_{\rm turb}$ is assumed to be 1000~km~s$^{-1}$ for all the grids.
In the current setting of $M_{\rm BH}=10^8M_\odot$, $10^5R_g/c$ corresponds to 1.66yr.
We calculated 9 angles $\times$ 3 snap shots $=$ 27 model energy spectra, and found that angle4 when time$=5\times10^5R_g/c$ reproduces the {\it XMM-Newton} data of PG 1211+143 best. Therefore we use this set in the main text.

\begin{table}
\centering
\caption{Wind properties with different snap shots}
\label{tab:compare}
\begin{tabular}{cccc}
\hline
\hline
Time ($\times10^5R_g/c)$& $\dot{M}_{\rm w}/\dot{M}_{\rm Edd}$ & $\dot{P}_{\rm w}/(L_{\rm Edd}/c)$ & $L_{\rm w}/L_{\rm Edd}$ \\
\hline
3 & 0.40 & 1.00 & 0.08\\
4 & 0.50 & 1.42 & 0.13\\
5 & 0.56 & 1.71 & 0.17\\
\hline
\end{tabular}
\end{table}

\section{Validity for introducing a pseudo-turbulent velocity}\label{sec:a2}
In this section we demonstrate that our method with a coarse grid gives similar results to the one with a finer grid.

The wind parameters (density, ionisation parameter, velocity, etc.) must be continuously changed unless a shock wave occurs. However in our Monte-Carlo calculation, we did not use grids as fine as the original hydro-simulation (Figure \ref{fig:wind_density} upper) due to the limited machine memory and calculation time.
The coarse grid we adopt (Figure \ref{fig:wind_density} lower) may produce some resolution artefact; some wind clumps may be artificially made and they will emphasise the discrete absorption features.
It is true that using finer grids is the ideal solution to avoid the resolution artefact, but unfortunately it is not a realistic one.
Instead of it, we introduced a pseudo-turbulent velocity. 

\subsection{Simulation for photoionised absorbers in the line of sight}
In order to reproduce the spectral synthesis, we assume that the wind material along the line of sight is described by the {\tt zxipcf} model in {\sc xspec}.
This model uses a grid of {\sc xstar} photoionised absorption.
Because it assumes a small turbulent velocity (200~km~s$^{-1}$),
we introduce the {\tt gsmooth} model, which smears spectral features with a Gaussian.
We note that this calculation is simplified compared with our Monte-Carlo simulation, but worthwhile to study the resolution artefact.

\begin{figure}
\centering
	\includegraphics[width=3.4in, angle=270]{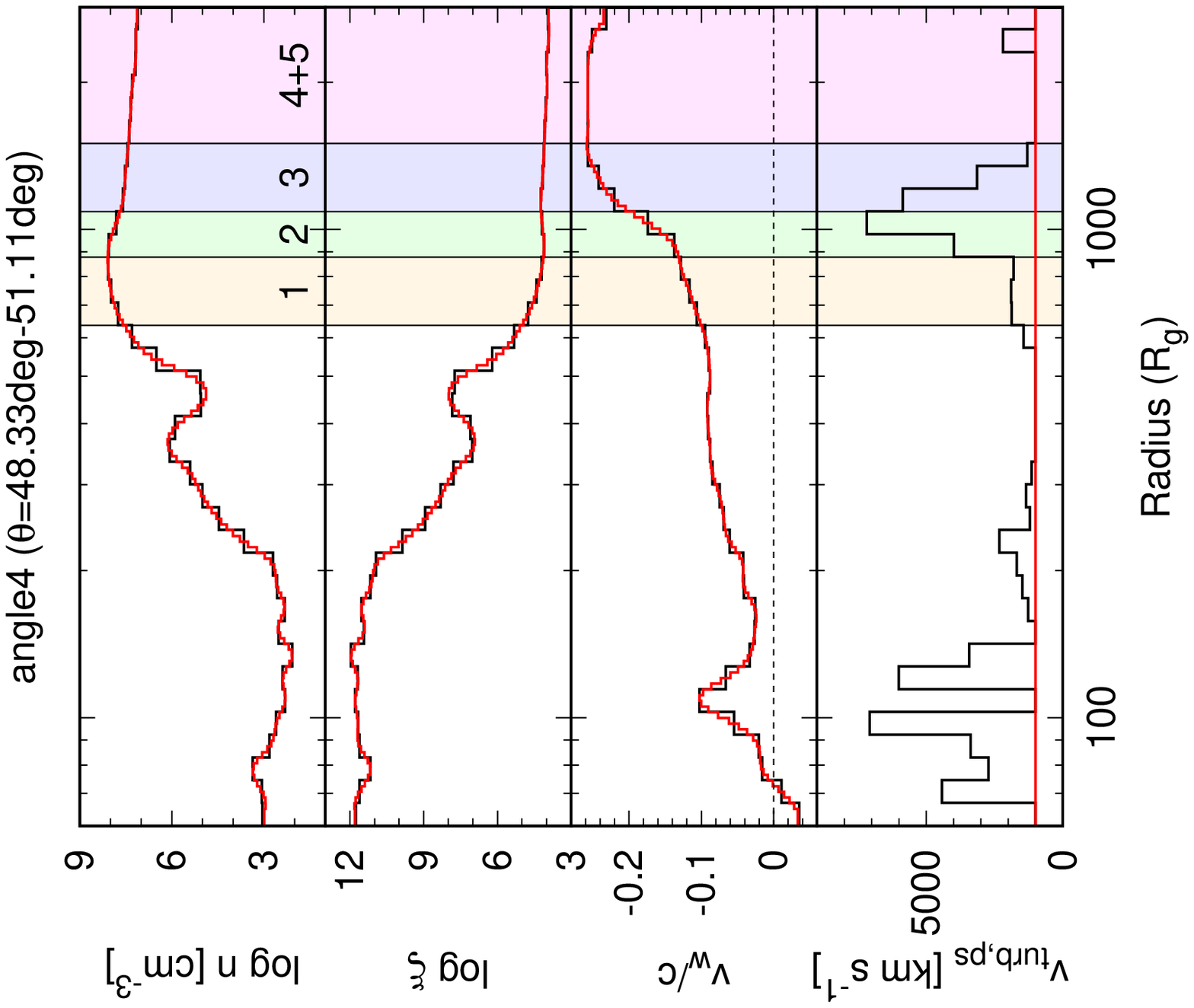}
    \caption{Radius dependence of the wind parameters (density, ionisation parameter, velocity, pseudo-turbulent velocity) along angle4.
    The black lines are the same as those in the Figure \ref{fig:tau_a4} in the main text.
    The red lines are the interpolated one, whose grids are 4 times as fine as the black ones. The pseudo-turbulent velocity in the red line is fixed as 1000~km~s$^{-1}$.}
    \label{fig:angle4new}
\end{figure}

\begin{figure}
\centering
	\includegraphics[width=3.2in, angle=270]{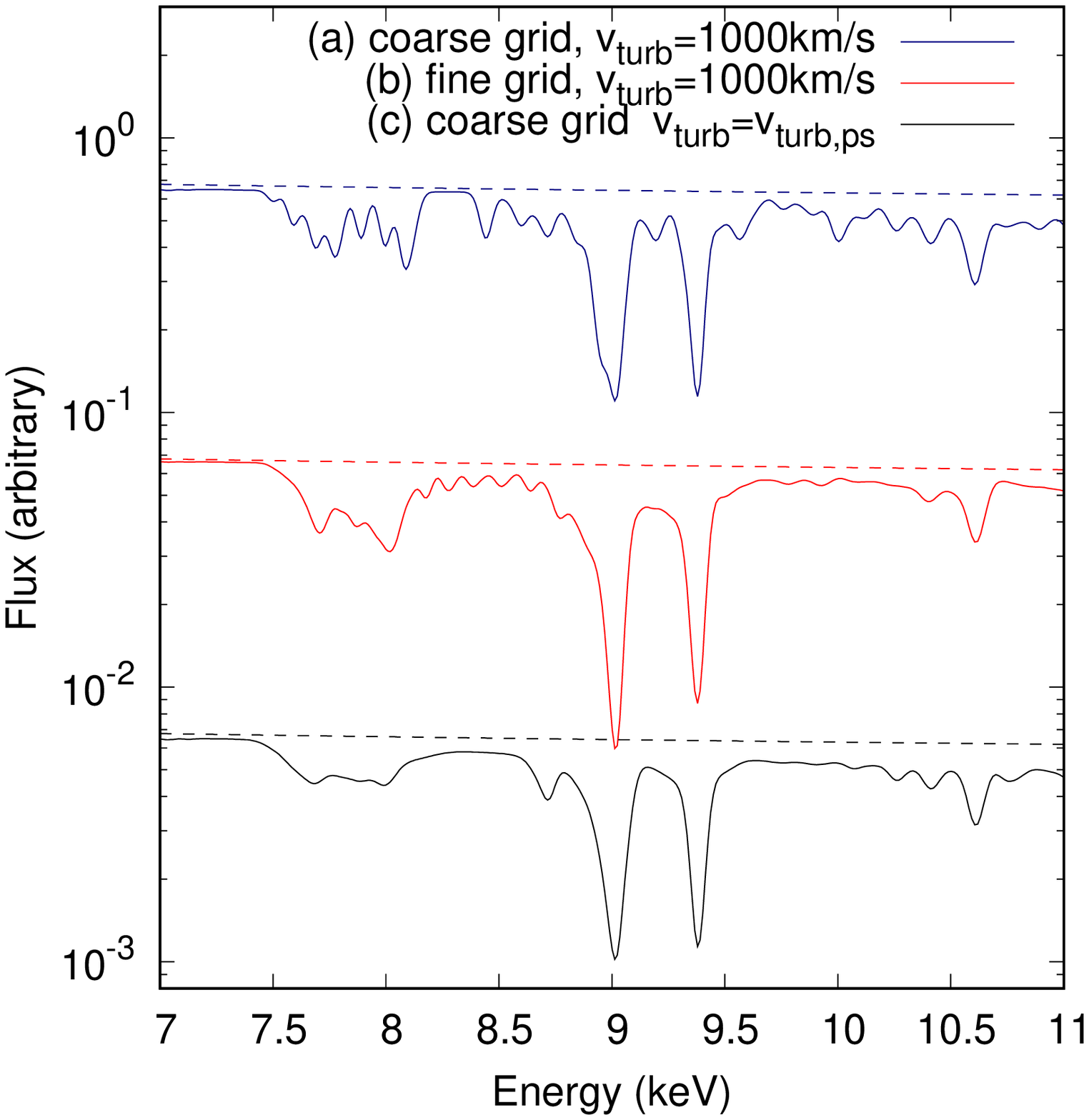}
    \caption{Spectral modelling for the wind along the line of sight. The dashed lines show the input power law.}
    \label{fig:interpolate}
\end{figure}

As a first step, we introduce 15 {\tt gsmooth$\times$zxipcf} on the power law with $\Gamma=2.2$, with the turbulent velocity of 1000~km~s$^{-1}$.
We use the absorbers within $650\,R_{\rm g}<r<3000\,R_{\rm g}$, i.e., the shaded region in the Figure \ref{fig:angle4new}.
This reproduces the coarse grid without the pseudo-turbulent velocity treatment (Figure \ref{fig:interpolate}(a)).
Complex and discrete features are seen. For example, the 7.5--8.2 keV band has 6 narrow lines. This is a resolution artefact as shown below.

As a next step, we interpolate the parameter change with 4 times finer grids (the red line in Figure \ref{fig:angle4new}), using the spline function.
This approximates a continuous change, and we introduce 60 {\tt gsmooth$\times$zxipcf}, with the turbulent velocity of 1000~km~s$^{-1}$.
Figure \ref{fig:interpolate}(b) shows the results. There are still small artifacts from the finite resolution elements, but these artefacts are much weaker than in (a). Interestingly we can clearly see that line blending, where the velocity shift between elements is smaller than the assumed turbulent width, gives two major K$\alpha$ absorption line systems, with slower and faster components at 7.5--8~keV and 8.5--9.5~keV.

As a last step, we compare the result with the coarse grid with the pseudo-turbulent velocity, which reproduces our simulation in the main text (Figure \ref{fig:interpolate}(c)).
Again, similar to the high resolution case (b), the spectral features are dominated by the two main velocity systems, though there are other broader absorption features also. 
This shows that our pseudo-turbulence approach can reproduce the finer grid case well.

\subsection{Bimodality of the wind}
\begin{figure}
\centering
	\includegraphics[width=2.2in,angle=270]{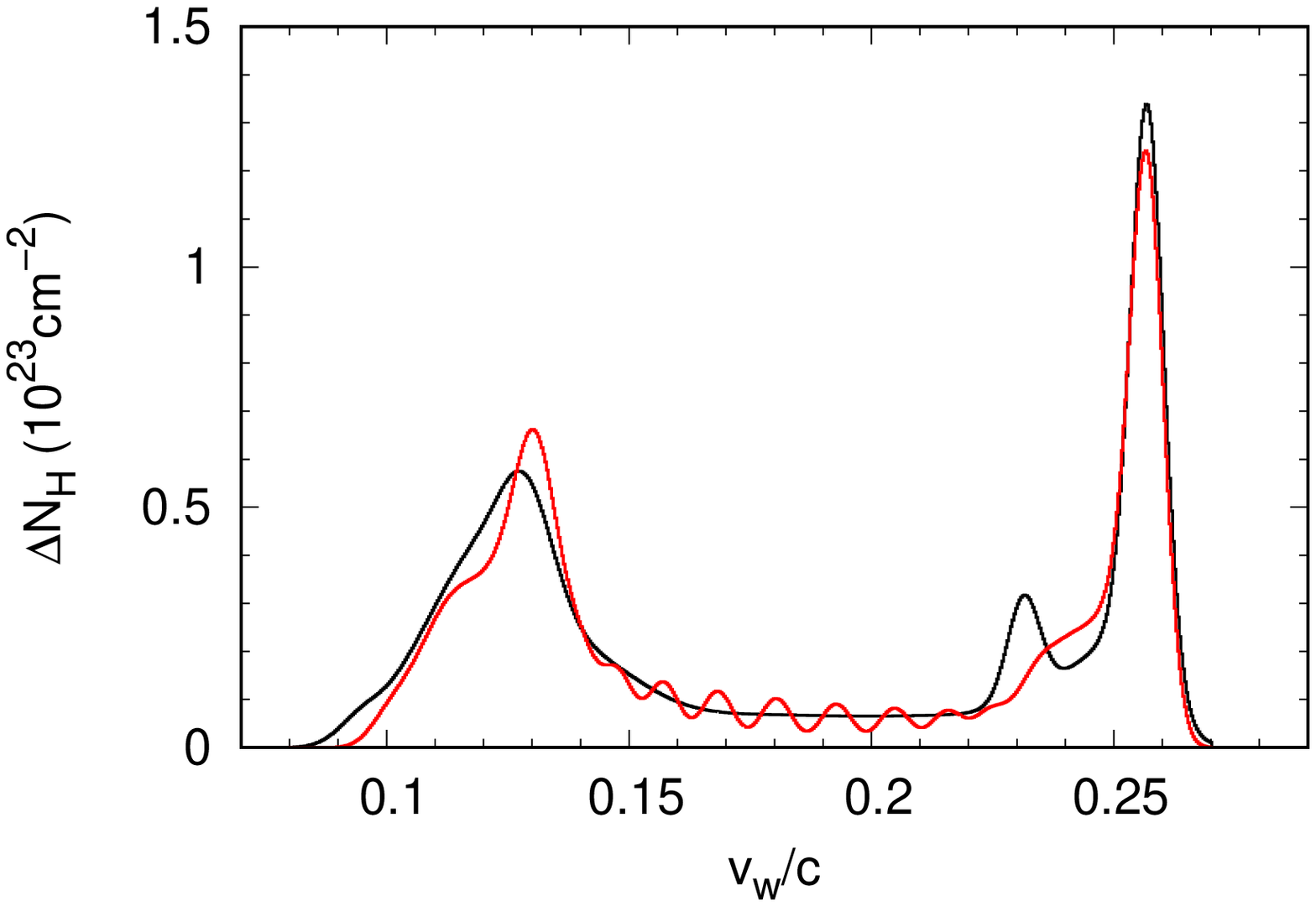}
    \caption{column distributions against the wind velocity. The data in $r>600\,R_g$ is used. The red line shows the fine grid case with $v_{\rm turb}=1000$~km~s$^{-1}$, whereas the black does the coarse grid case with $v_{\rm turb}=v_{\rm turb,ps}$.}
    \label{fig:hist}
\end{figure}

We also check the bimodality of the wind with a more straightforward manner.
Figure \ref{fig:hist} shows the column distributions against the wind velocity in the line formation range ($r>600\,R_g$).
We assumed that each bin in Figure \ref{fig:angle4new} has a Gaussian profile with $\sigma_v=v_{\rm turb}$ and calculated the total distribution as a summation of each Gaussian.
The colour is same as Figures \ref{fig:angle4new} and \ref{fig:interpolate};
the red line is calculated from the fine grid with $v_{\rm turb}=1000$~km~s$^{-1}$, whereas the black is from the coarse grid with $v_{\rm turb}=v_{\rm turb,ps}$.
They are very similar to each other and have a bimodal distribution, with the slower and faster peaks.
This shows that the discrete absorption features are a physical feature of the UV line driven disc wind simulation, and are present  no matter which method we adopt. We note that UV line driven disc winds generically show variability in total column density with time, and since these are connected to a wind outflow, this also means that the density at a given position changes with time (see \citealt{sch09}).

\bsp	
\label{lastpage}
\end{document}